\documentclass[aps,prl,superscriptaddress,twocolumn]{revtex4}

\usepackage{graphicx}
\usepackage{dcolumn}
\usepackage{bm}
\usepackage{color}
\usepackage[table]{xcolor}
\usepackage{amsmath}
\usepackage{amssymb}
\usepackage[caption=false]{subfig}
\usepackage{amsmath}
\usepackage{float}
\usepackage{natmove}
\usepackage{multirow}
\usepackage{setspace}
\usepackage{array}
\usepackage{booktabs}
\usepackage{rotating}
\usepackage{comment}
\usepackage{xr-hyper}
\makeatletter
\newcommand*{\addFileDependency}[1]{
  \typeout{(#1)}
  \@addtofilelist{#1}
  \IfFileExists{#1}{}{\typeout{No file #1.}}
}
\makeatother

\usepackage[colorlinks=true,linkcolor=blue,citecolor=blue,urlcolor=blue,filecolor=blue]{hyperref}
\usepackage{physics}
\usepackage{mathrsfs}
\usepackage{tabularx}
\usepackage{layouts}
\usepackage{ulem}
\setcitestyle{numbers,square}

\newcolumntype{C}{>{\centering\arraybackslash}X}

\definecolor{tangerine}{rgb}{0.944,0.522,0}
\definecolor{brown}{rgb}{0.633,0.156,0.156}
\definecolor{lime}{rgb}{0.5,1.0,0.0313}
\definecolor{limedark}{rgb}{0.333, 0.666, 0.020}
\definecolor{applegreen}{rgb}{0.55, 0.71, 0.0}
\definecolor{green1}{rgb}{0.0, 0.5, 0.0}
\definecolor{green2}{rgb}{0.25, 0.5, 0.016}
\definecolor{BluBondi}{rgb}{0.00,0.58,0.71}
\definecolor{myred}{rgb}{0.784, 0.063, 0.180}  
\definecolor{mygreen}{rgb}{0.478,0.604,0.004}  
\definecolor{myblue}{rgb}{0.059,0.298,0.506}   

\newcommand{\editor}[2]{%
  \expandafter\newcommand\csname #1note\endcsname[1]{%
    \textcolor{#2}{(\textbf{#1:} {\it ##1})}}%
  \expandafter\newcommand\csname #1\endcsname[1]{%
    \textcolor{#2}{##1}}%
  \expandafter\newcommand\csname #1cancel\endcsname[1]{%
    \textcolor{#2}{\sout{##1}}}%
  \expandafter\newcommand\csname #1change\endcsname[2]{%
    \textcolor{#2}{\sout{##1} ##2}}%
  \newenvironment{#1text}{\color{#2}}{\color{black}}
}

\editor{AF}{limedark}
\editor{NM}{green}
\editor{TC}{red}

\makeatletter
\def\maketitle{
\@author@finish
\title@column\titleblock@produce
\suppressfloats[t]}
\makeatother

\begin{document}
\title{Dynamical Hubbard approach to correlated materials: \\
the case of transition-metal monoxides}

    \newcommand{\epfl}{Theory and Simulation of Materials (THEOS), and National Centre for Computational Design and Discovery of Novel Materials (MARVEL), \'Ecole Polytechnique F\'ed\'erale de Lausanne, CH-1015 Lausanne, Switzerland}

    \newcommand{\piesai}{PSI Center for Scientific Computing, Theory and Data, 5232 Villigen PSI, Switzerland}   
       
    \newcommand{\etsf}{European Theoretical Spectroscopy Facility (ETSF)}

    \newcommand{\champittet}{Colleege Champittet, Pully, Switzerland}

    \newcommand{\caltech}{Department of Applied Physics and Materials Science, California Institute of Technology, Pasadena, California 91125, USA}

	\author{Mario Caserta\footnote{These two authors contributed equally.}}
	\email[]{mario.caserta@epfl.ch}
	\affiliation{\epfl}

      \author{Tommaso Chiarotti$^*$}
	\affiliation{\caltech}

    \author{Marco Vanzini}
	\affiliation{\champittet}\affiliation{\etsf}
    
	    \author{Nicola Marzari}
	\affiliation{\epfl}\affiliation{\piesai}
%
%
\date{\today}

%
%
%
\begin{abstract}
   Electronic correlations beyond static mean-field theories are of fundamental importance in describing the properties of complex materials - such as transition-metal oxides - where the low-energy physics is driven by localized $d$ or $f$ electrons. 
   Here, we show that it is possible to capture these correlations with a local and dynamical self energy, extending to the spin-polarized and multi-site case our recently introduced dynamical Hubbard functional formulation.
   We apply this formalism to the prototypical transition-metal monoxide series of MnO, FeO, CoO, and NiO in their ground state, finding excellent agreement with experiments for the spectral properties. The results are comparable or improved with respect to state-of-the-art theories, both for the densities of states and for the spectral functions---including band renormalization and spectral weight transfer---in a numerically efficient and physically transparent treatment of correlations amenable to the study of realistic, complex materials.
\end{abstract}

%
%
\maketitle
%

Binary transition-metal oxides MnO, FeO, CoO and NiO are some of the most studied and characterized materials, both experimentally~\cite{sawatzky_magnitude_1984,van_elp_electronic_1991_MnO,zimmermann_electronic_1999,shen_photoemission_1990,shen_electronic_1991,kuhlenbeck_molecular_1991} and theoretically~\cite{kunes_nio_2007,mandal_influence_2019,mandal_systematic_2019,abdallah_quasiparticle_2024,zhang_symmetry-breaking_2020,cococcioni_linear_2005}. 
These compounds have been categorized respectively as Mott and charge-transfer insulators, within the common Zaanen-Sawatzky-Allen scheme \cite{zaanen_band_1985} and they are insulating both above and below their N\'{e}el temperature.
They represent ideal test cases for electronic structure methods thanks to both their simple crystal structures (rock-salt (Fm3m) in the paramagnetic phase and slightly distorted rhombohedral (R3m) with two TM ions per the unit cell in the low-temperature AFM-II phase), and for displaying gradually varying properties throughout the filling of the $3d$ shell.
Describing quantitatively the complex interplay between the electronic localization on the transition-metal (TM) sites and the hybridization with the ligands is a challenging task for Kohn-Sham density-functional theory (DFT) with standard semi-local functionals~\cite{anisimov_band_1991,shen_electronic_1991,mandal_systematic_2019}, as it underestimates the electronic gap in MnO and NiO, and fails to predict the insulating state of CoO and FeO. 
These shortcomings are inherited by many-body perturbation theory (MBPT) approaches, such as G$_0$W$_0$ and GW$_0$, when they start from the DFT metallic electronic structure~\cite{mandal_systematic_2019,aryasetiawan_electronic_1995}. 
Improvements in the description of the excitation spectrum can be obtained using, e.g., DFT+U and hybrid functionals~\cite{anisimov_band_1991,anisimov_first-principles_1997,cococcioni_linear_2005,mandal_systematic_2019}, where the insulating states and gap magnitudes are recovered, gauged by the effective $U$ and exchange-fraction parameters. These are often fitted to reproduce experimental data, thus hindering the predictive first-principles nature of the methods, or are unable to systematically reproduce the full spectral properties across the series~\cite{cococcioni_linear_2005,anisimov_first-principles_1997,mandal_systematic_2019}. 
To improve on G$_0$W$_0$ and GW$_0$ results, one can instead either start from DFT+U~\cite{jiang_first-principles_2010,das_convergence_2015} or hybrid ground-states~\cite{rodl_quasiparticle_2009}, and/or introduce vertex corrections in quasi-particle self-consistent extensions (QSG\^W) \cite{abdallah_quasiparticle_2024,cunningham_qs_2023}.
While this improves the description of gap magnitudes and hybridization effects typical of charge-transfer insulators, the ability to describe $k$-resolved spectral weight transfer and bands renormalization, prototypical features of electronic correlations~\cite{tomczak_asymmetry_2014,boehnke_when_2016}, requires a theoretical framework that incorporates complex and frequency-dependent self-energies, giving access to correlation-induced particle lifetimes~\cite{mandal_influence_2019,mandal_systematic_2019}.
DFT+U and hybrids are inherently static mean-field solutions, while in full GW or QSGW and QSG\^W the full frequency dependence is commonly discarded because of the heavy computational cost and numerical complexity.
Given that dynamical frameworks allow for energetic corrections and localization pathways inaccessible to static solutions, DFT+DMFT~\cite{held_realistic_2006,kotliar_electronic_2006} has become the theoretical tool to model complex transition-metal oxides~\cite{mandal_systematic_2019,paul_applications_2019}.
In DMFT, the local correlated subspace, often the TM $d$ or $f$ shell, is mapped onto an effective Anderson impurity model~\cite{anderson_localized_1961} which can be solved exactly, often with Monte Carlo methods~\cite{gull_continuous-time_2011}, and embedded back into the DFT solution in a self-consistent cycle. Thus, DFT+DMFT allows to describe local dynamical electronic correlations and fluctuating magnetic moments, 
improving the description of electronic gaps, bands renormalization and spectral weight transfer. Though, it also leads to high computational costs and increased numerical complexity, with exact solutions scaling exponentially with system size~\cite{werner5,gull_continuous-time_2011}. 

In this Letter, we show that a simple and transparent dynamical formulation, obtained from the generalization of DFT+U to host a local frequency-dependent screened interaction, is able to capture all the correlated-electrons signatures of gap opening, bands renormalization, and spectral weight transfer, in excellent agreement with experimental PES/IPES data~\cite{shen_electronic_1991,van_elp_electronic_CoO,van_elp_electronic_1991_MnO,kuhlenbeck_molecular_1991} and state-of-the-art DFT+DMFT result~\cite{mandal_influence_2019,mandal_systematic_2019}.
This is achieved by solving the Dyson equation with the algorithmic-inversion method on sum over poles (AIM-SOP) for a dynamical Hubbard functional (dynH)~\cite{chiarotti_energies_2024}, generalized here to treat magnetic systems with multiple TM metal sites. 
The functional derivative of the dynH functional with respect to the local Green's function provides the dynH self-energy localized on the correlated manifold.
We apply here the framework to study the MnO, FeO, CoO and NiO series, obtaining beyond state-of-the art results for the spectral properties of these materials.
In all cases a very good agreement with experiments is found, including the small gap for FeO, in agreement with the experimental results~\cite{zimmermann_electronic_1999} and in contrast with DFT+DMFT~\cite{mandal_influence_2019} and QSG\^W~\cite{abdallah_quasiparticle_2024} that predict a wide gap for all four monoxides. 
Moreover, it is shown that in the case of NiO the present approach predicts electronic bands present in the experimental ARPES data~\cite{shen_electronic_1991,kuhlenbeck_molecular_1991} that disappear in the strong correlation regime of the DMFT impurity solution ~\cite{kunes_nio_2007,mandal_influence_2019}.

The dynamical Hubbard functional (dynH)~\cite{chiarotti_energies_2024} has been introduced to address local dynamical correlations, giving accurate results for the spectral properties of prototypical correlated oxide SrVO$_3$, in agreement with experimental data~\cite{backes_hubbard_2016} and state-of-the-art GW+EDMFT calculations~\cite{boehnke_when_2016,petocchi_screening_2020}.
Here, we generalize the functional to the multi-site collinear spin-polarized case, in order to treat magnetic systems. The generalization of Eq.~(8) in~\cite{chiarotti_energies_2024} to spin-dependent cases involves the spin-polarization of the momentum-bands Green's function and of the RPA screened interaction. From the convolution of the two, one arrives at a local spin-dependent exchange and correlation self-energy of the form: 
\begin{equation}
\begin{split}
    \mathbf{\Sigma}_{dynH} ^{\sigma,I}(\omega) = & 2\pi i \frac{\partial \Phi_{dynH}[\mathbf{G}_I^{\sigma}]}{\partial \mathbf{G}_I^{\sigma}} \\
    = & - \int d\omega' U_I(\omega')\mathbf{G}_I^{\sigma}(\omega + \omega')  +  \frac{U_{I}^{\infty}}{2}.
\end{split}    
\label{eq:dynHUB_SE}
\end{equation}
The local spin Green's function is defined as $\mathbf{G} = P^{\dagger} G P $, where the projectors $P$ transform from the momentum-band ($\mathbf{k},n$) indexes  to the ($m$) index, representing the magnetic quantum number of an atomic-like orbitals $\phi_m$ (e.g. $3d$ shell with $l=2$ and $m=-2,-1,0,1,2$) centered on each transition-metal site $I$ in the cell. $U(\omega)$ is the local average of the direct matrix elements in the basis of the $N_d$ atomic-like orbitals~\footnote{The spin dependence of the local basis is related to spin dependence of the Wannier functions used as projectors in the $U(\omega)$ calculation, to the ortho-atomic projectors used in (\ref{eq:dynHUB_SE}) which bear no spin dependency. The MLWF~\cite{marzari_maximally_2012} built from the TM $3d$, $3p$, $4s$ and O $2p$ orbitals } of the screened screened Coulomb interaction \cite{golze_gw_2019}
 \begin{equation}
     U_I(\omega) = \frac{1}{N_d}\sum_{\substack{mm' \\ \sigma \sigma'}}  \bra{\phi_{m,I}^{\sigma} \phi_{m',I}^{\sigma'}} W_{RPA}(\mathbf{r},\mathbf{r'},\omega)\ket{\phi_{m,I}^{\sigma}\phi_{m',I}^{\sigma'}},
 \end{equation} 
calculated from the RPA spin-DFT polarization function, and $U^{\infty}$ is its fully unscreened value (local matrix element of bare Coulomb potential).
\begin{figure}  
    \centering
    \includegraphics[width=0.45\textwidth]{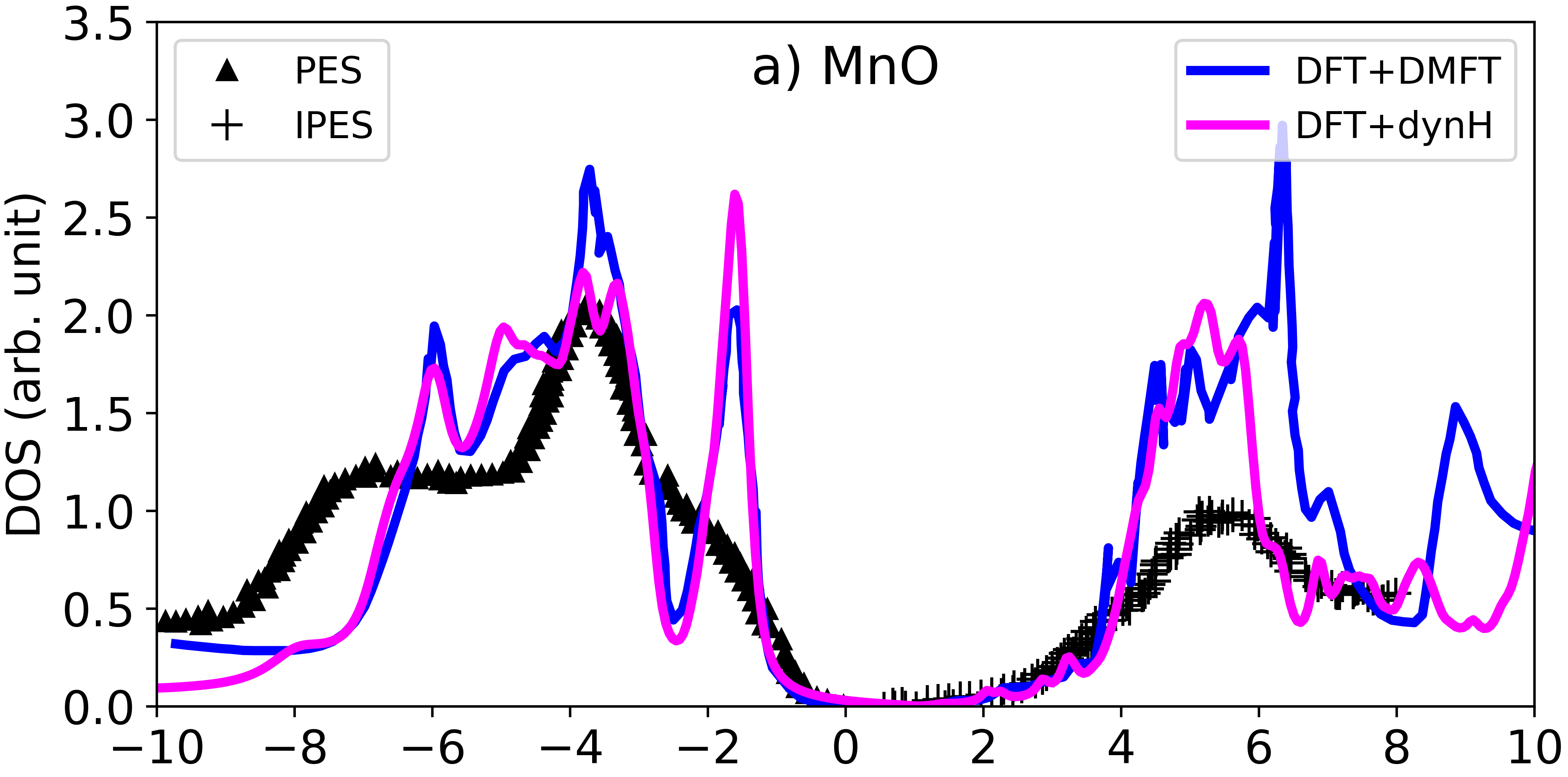}   \includegraphics[width=0.45\textwidth]{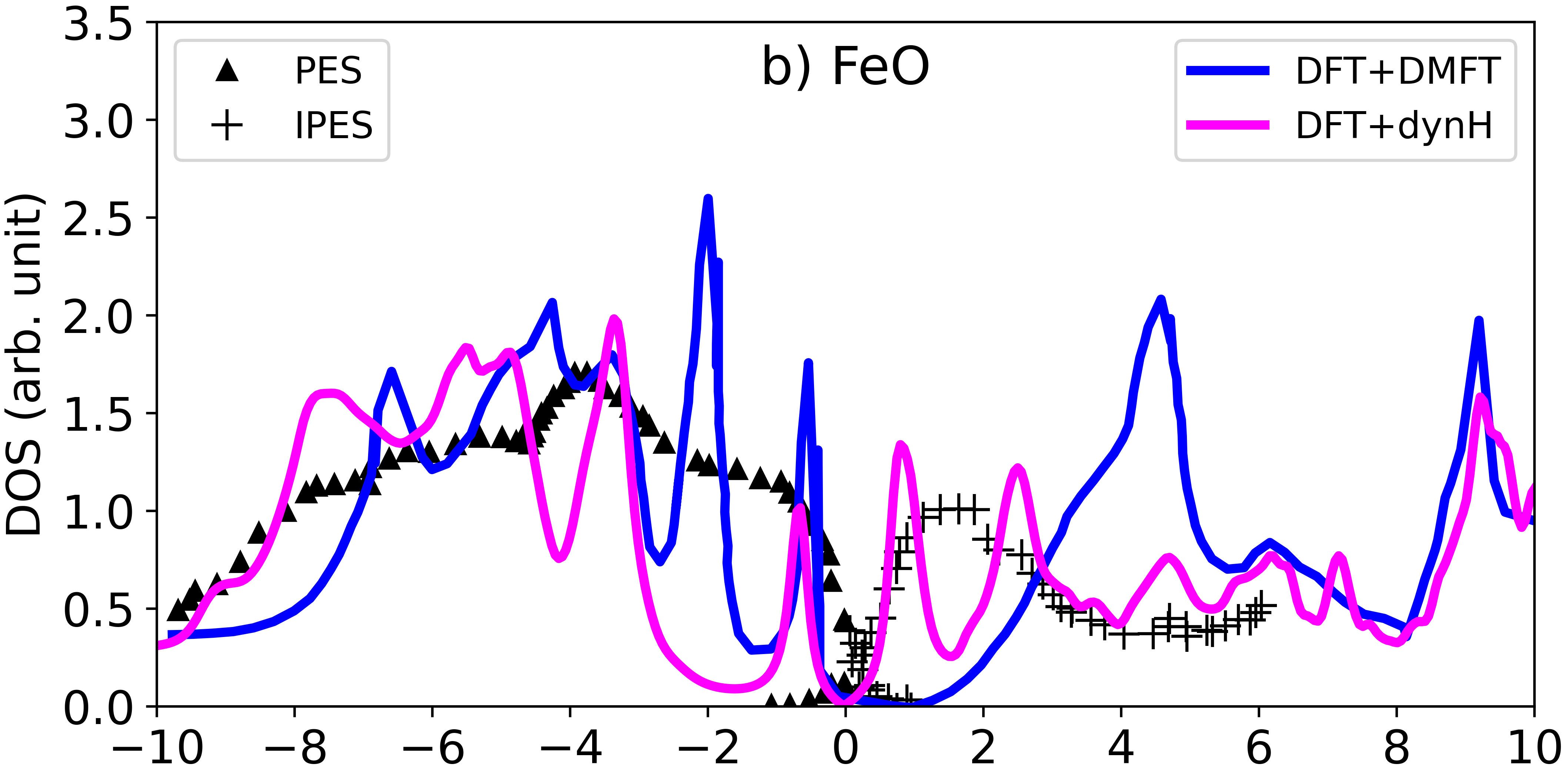}
    \includegraphics[width=0.45\textwidth]{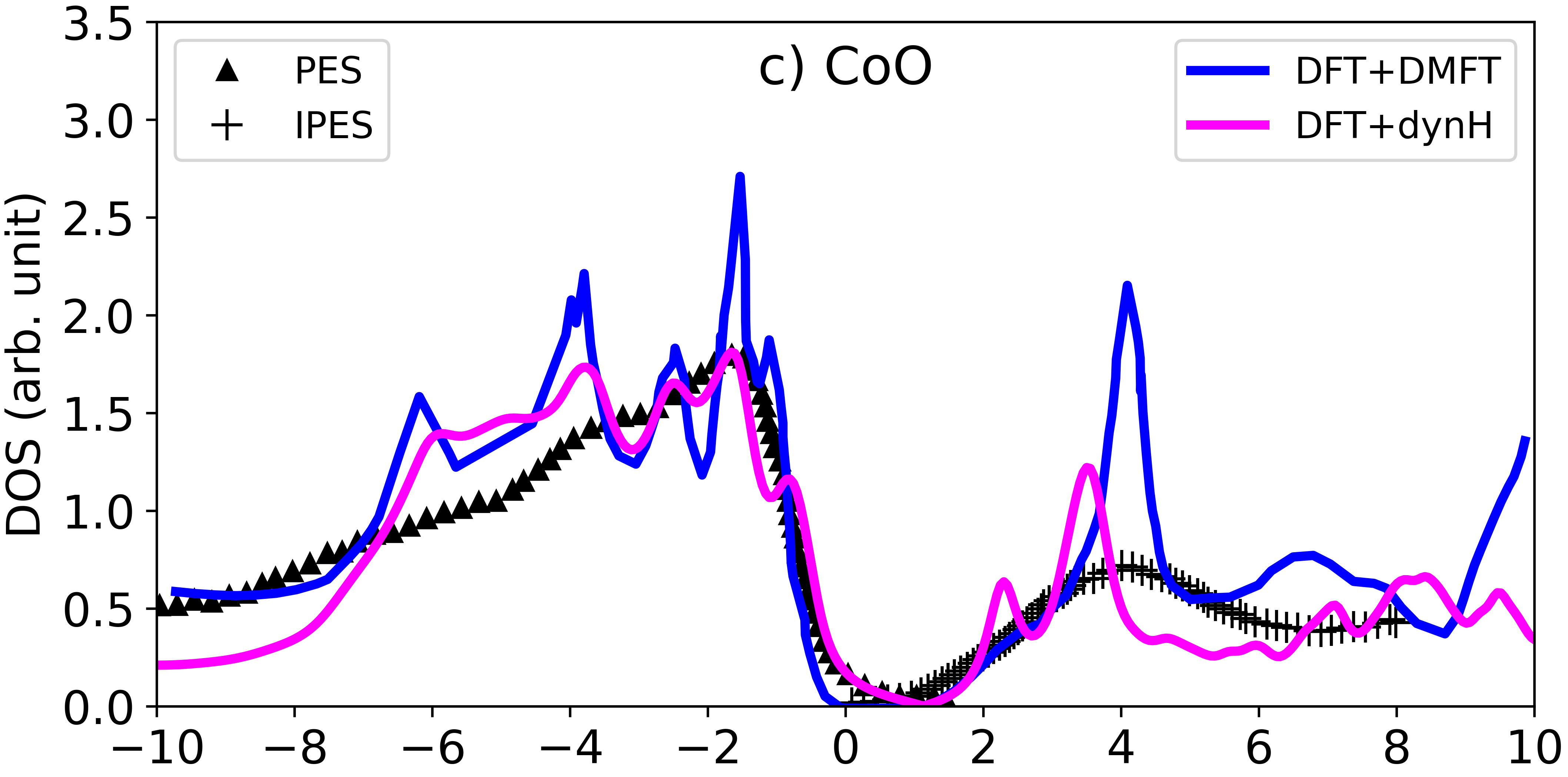}    
    \includegraphics[width=0.45\textwidth]{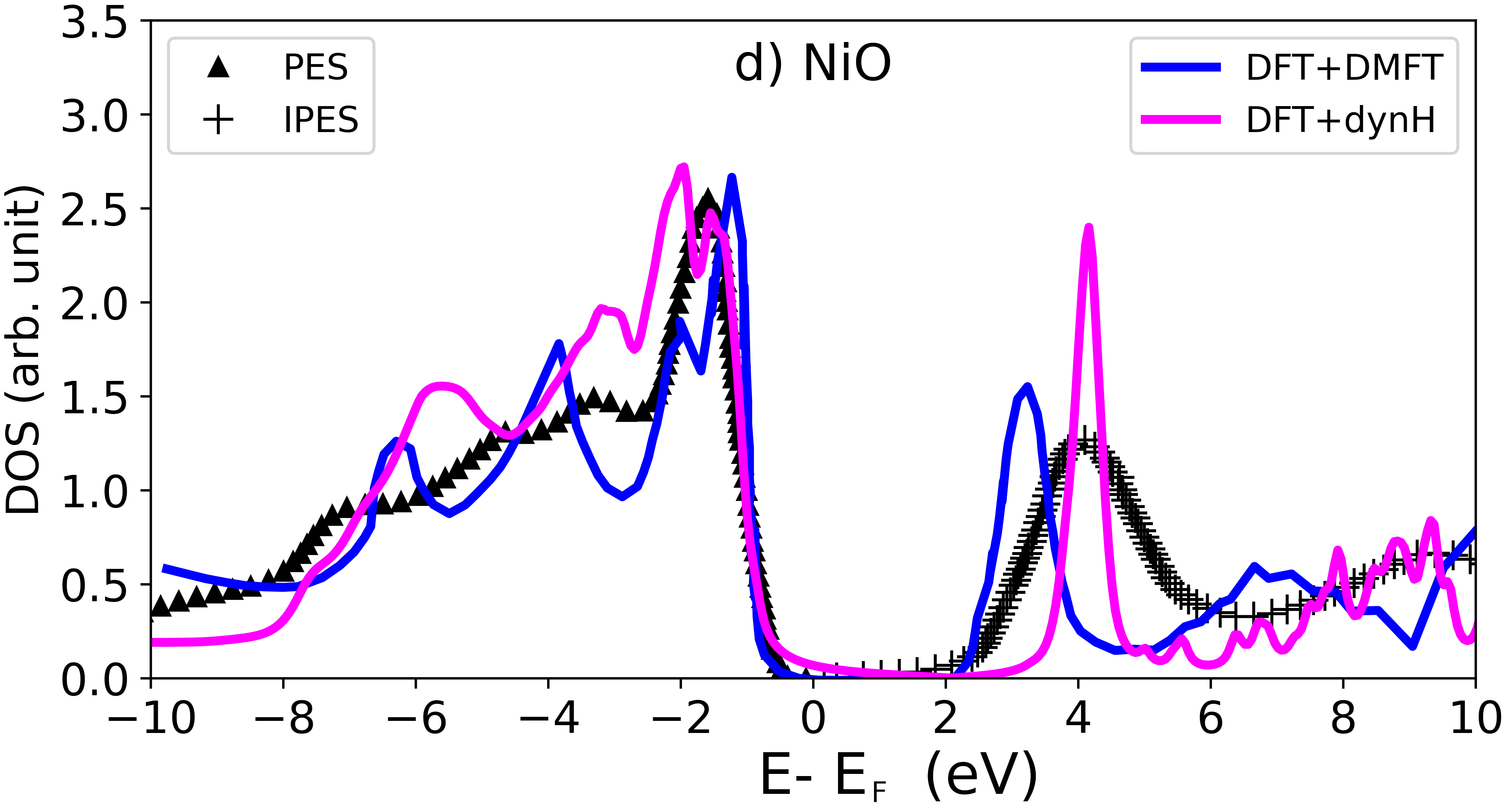}   
    \caption{Density of states (DOS) for one-shot DFT+dynH (solid magenta lines) for MnO (a), FeO (b), CoO (c) and NiO (d), compared to DFT+DMFT~\cite{mandal_influence_2019} (solid blue lines).   
    Experimental data in black markers from~\cite{van_elp_electronic_1991_MnO} for MnO,~\cite{zimmermann_electronic_1999} for FeO,~\cite{van_elp_electronic_CoO} for CoO and~\cite{hufner_optical_1992} for NiO. A broadeing of 0.1 eV has been used.}
    \label{dosses}
\end{figure}
The expression in Eq.~\eqref{eq:dynHUB_SE} can be seen as a complex and frequency-dependent generalization of the Dudarev rotationally invariant formulation of DFT+U potential~\cite{dudarev_electron-energy-loss_1998}, now hosting a dynamical and complex time-ordered effective interaction potential, plus a static double-counting term \cite{chiarotti_energies_2024,vanzini_towards_2023}. It also extends the model self-energy derived by Vanzini and Marzari~\cite{vanzini_towards_2023} to a full functional formulation. It is important to stress that the local dynamical interaction used here is screened by all, local and non-local, processes at the RPA level. 
The local self-energy of Eq.~\eqref{eq:dynHUB_SE} is then upfolded to the $k$ space, using $\Sigma = P \mathbf{\Sigma}_{dynH} P^{\dagger}$.
The $k$-resolved spectral function and density of states are then obtained from the Green's function that solves the Dyson equation $G^{-1} = G_0^{-1} - \Sigma$. 
The solution of the Dyson equation is achieved via the aforementioned algorithmic inversion method on sum over poles (AIM-SOP)~\cite{chiarotti_unified_2022,chiarotti_energies_2024}, where both the local screened Coulomb interaction and local Green's function are represented as sums over first-order poles and amplitudes. 
Importantly, when looking at spectral properties, AIM-SOP avoids the use of imaginary Matsubara frequencies as in the most common impurity solutions used in DMFT~\cite{gull_continuous-time_2011}, which requires the analytic continuation to the real axis thus hindering the accuracy of the calculations~\cite{silver_maximum-entropy_1990}. 
We refer to the Supplemental Material (SM) for the SOP representations of the local dynamical screened $U(\omega)$ and to Ref.~\cite{chiarotti_energies_2024} for the overall AIM-SOP framework. 
Importantly, we find that the full solution of the Dyson equation for these materials is crucial when looking at $k$-resolved spectral functions, where the description of band crossings has to be properly accounted for.
\begin{figure}       \includegraphics[width=0.45\textwidth]{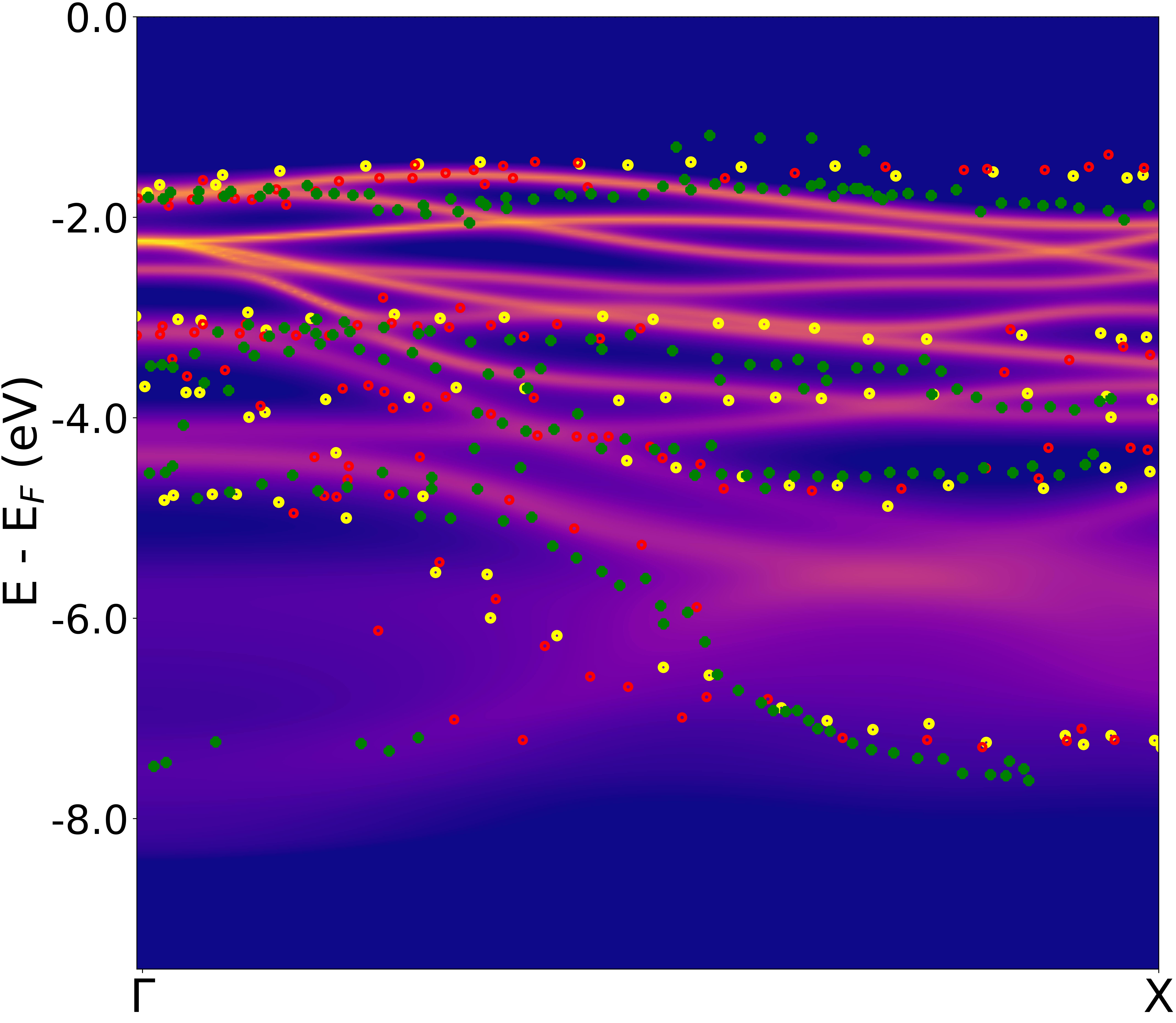}    
     \caption{DFT+dynH spectral function (purple/blue colormap) and experimental ARPES data for NiO (AFM). 
     The red (normal) and yellow (off-normal) dots represent emission data from \cite{shen_electronic_1991}.
     Green squares data are normal emission from \cite{kuhlenbeck_molecular_1991}. Experimental data have been aligned at the top of the valence band in the $\Gamma$ point, being it the most accurate in resolution \cite{shen_electronic_1991}. A broadening of 0.015 eV has been used.}  
    \label{NiO_arpes_dft}
\end{figure}
For the DFT ground-state calculation, we use the r2SCAN functional~\cite{furness_accurate_2020}, owing to its correct prediction of an insulating phase in all four compound~\cite{zhang_symmetry-breaking_2020} on an equal footing. 
Computational details, DFT band structures, DFT+dynH projected density of states can be found in the (SM).

In Fig.~\ref{dosses} we show the total density of states of one-shot DFT+dynH for the AFM phase of the four monoxides, together with state-of-the-art charge self-consistent DFT+eDMFT~\cite{mandal_influence_2019} and PES/BIS experiments~\cite{van_elp_electronic_1991_MnO,sawatzky_magnitude_1984,van_elp_electronic_CoO,zimmermann_electronic_1999}.
For MnO, Fig.~\ref{dosses}(a), the occupied DOS is consistent with DFT+DMFT results~\cite{mandal_influence_2019}, where the overall bandwidth is slightly contracted with respect to the experimental one~\cite{van_elp_electronic_1991_MnO}, as the high-energy satellites above remain elusive to both DFT+DMFT~\cite{mandal_influence_2019}, DFT+dynH (and also in QSG\^{W} calculations~\cite{abdallah_quasiparticle_2024}). The LUMO excitation at 5 eV above Fermi appears to be in agreement with IPES data in position, width and intensity. 
For FeO, Fig.~\ref{dosses}~(b), DFT+dynH predicts a small gap of circa 2 eV, in agreement with the IPES data in Ref.~\cite{zimmermann_electronic_1999} and in contrast with state-the-art DFT+DMFT~\cite{mandal_systematic_2019} and QSQ\^W\cite{abdallah_quasiparticle_2024}, that predict wide gaps for all four monoxide. The decrease in intensity after the first excitations is also wel described by DFT+dynH.
It can be seen that the main difference between DFT+dynH and the DFT+DMFT results in the occupied part is the larger distance between the $a_{1g}$ HOMO excitation and the rest of the occupied spectrum. The O($2p$)/Fe($3d$) peak of $e_g$ symmetry in fact appears around -4 eV, in contrast with the -2.4 eV of DFT+DMFT, as it can be seen also in the $k$-resolved spectral functions, Panel (b) of Fig.~\ref{spectral_functions} and Fig. 1 of \cite{mandal_influence_2019}. The dip between the first and second ionization peaks (appearing also in QSG\^W \cite{abdallah_quasiparticle_2024}) can't be seen in the PES spectrum, though the total valence bandwidths of circa 10 eV in DFT+dynH match quite well. 
For CoO, in Fig.~\ref{dosses}~(c), the DOS compares well with both DFT+DMFT and PES data. The main difference with the former resides in the first addition peaks, since in DFT+dynH the two empty states of $t_g$ and $e_g$ symmetry have comparable weight, whereas in DFT+DMFT the weight is mainly transferred to the higher energy one, resulting in a single peak and a minor bump at 2.4 eV. The main peak at 3.5 eV is slightly lower than the DFT+DMFT one at 4 eV. 
For NiO, Fig.~\ref{dosses}(d), it can be seen that the agreement between the main experimental peaks~\cite{hufner_optical_1992}, in both addition and removal excitations, is very good, especially for the main addition peak at 4 eV and the sharp increase in the occupied spectrum.
In Fig.~\ref{NiO_arpes_dft}, NiO ARPES data and $k$-resolved spectral functions along the $\Gamma X$ direction are shown, where all data from normal~\cite{kuhlenbeck_molecular_1991,shen_photoemission_1990} and off-normal~\cite{shen_electronic_1991} emission are collected. 
The overall DFT+dynH bandwidth of 6 eV is consistent with the ARPES data. The flat Ni($3d$)/O($2p$) bands at the top, associated with the Zhang-Rice doublet bound state~\cite{bala_zhang-rice_1994,zhang_effective_1988}, can be recognized, as well as the flat Ni($3d$)/O($2p$) flat bands around -3 eV. 
The two dispersing O($2p$) bands departing from $\Gamma$ and extending at around -4.4 eV and -8 eV are also present. 
As it can be seen from the projected DOS (see also Fig.~S6 SM), the actual character of the valence band excitations is mixed, confirming the charge-transfer nature of the compound, as the first addition peak is mainly of Ni $e_g$ character. 
\begin{figure}
    \includegraphics[width=0.40\textwidth]{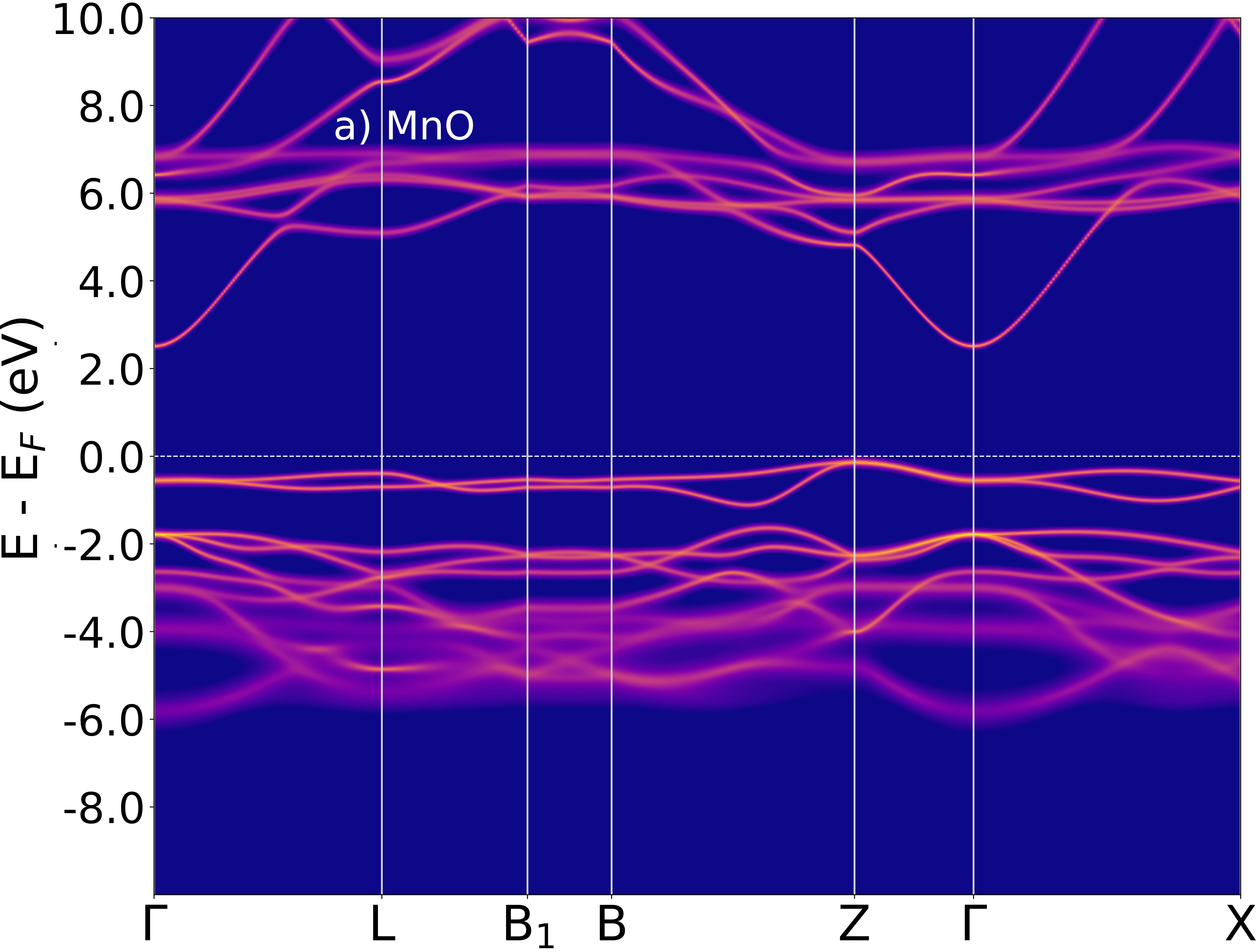}
   \includegraphics[width=0.40\textwidth]{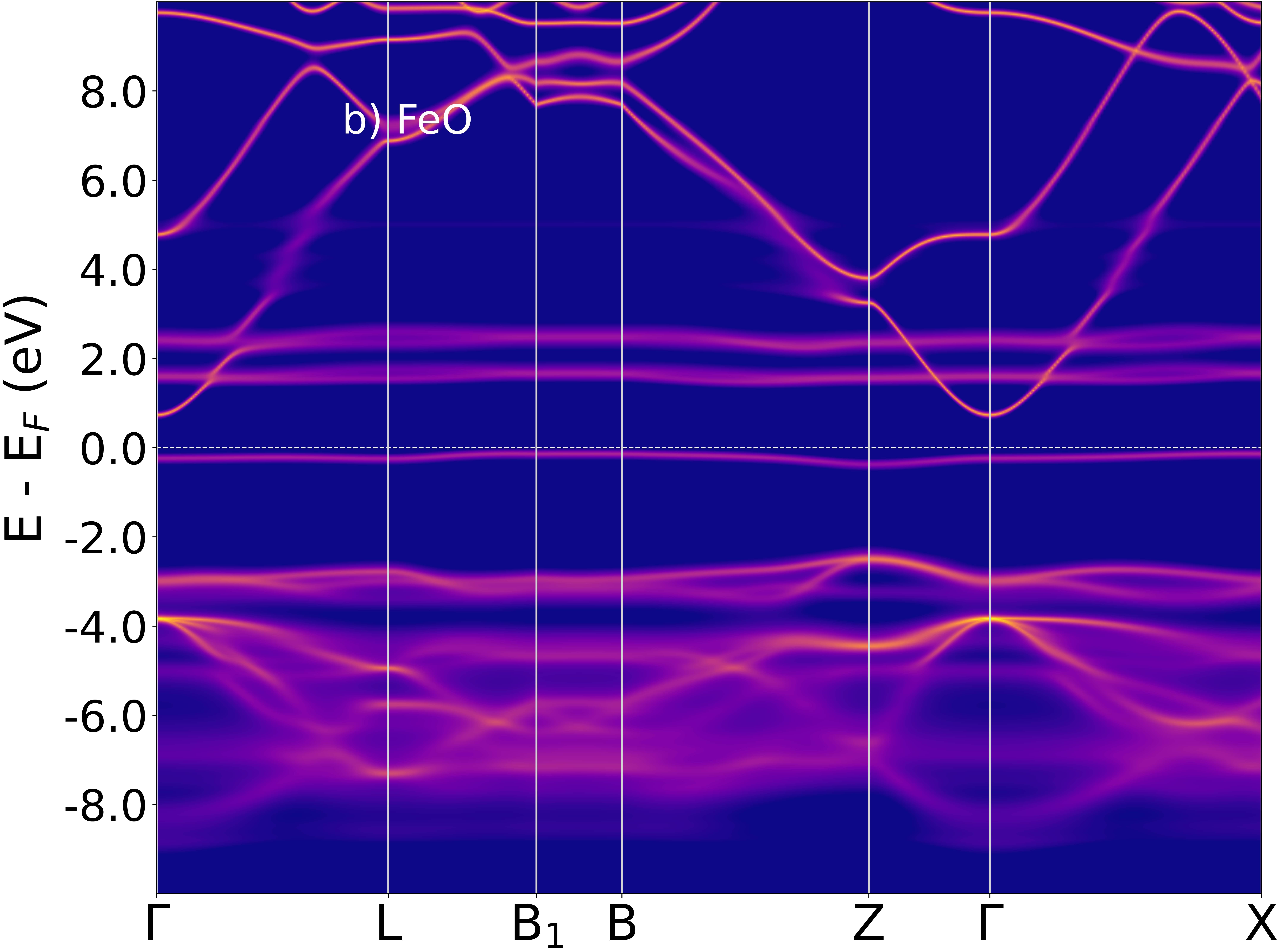}
    \includegraphics[width=0.40\textwidth]{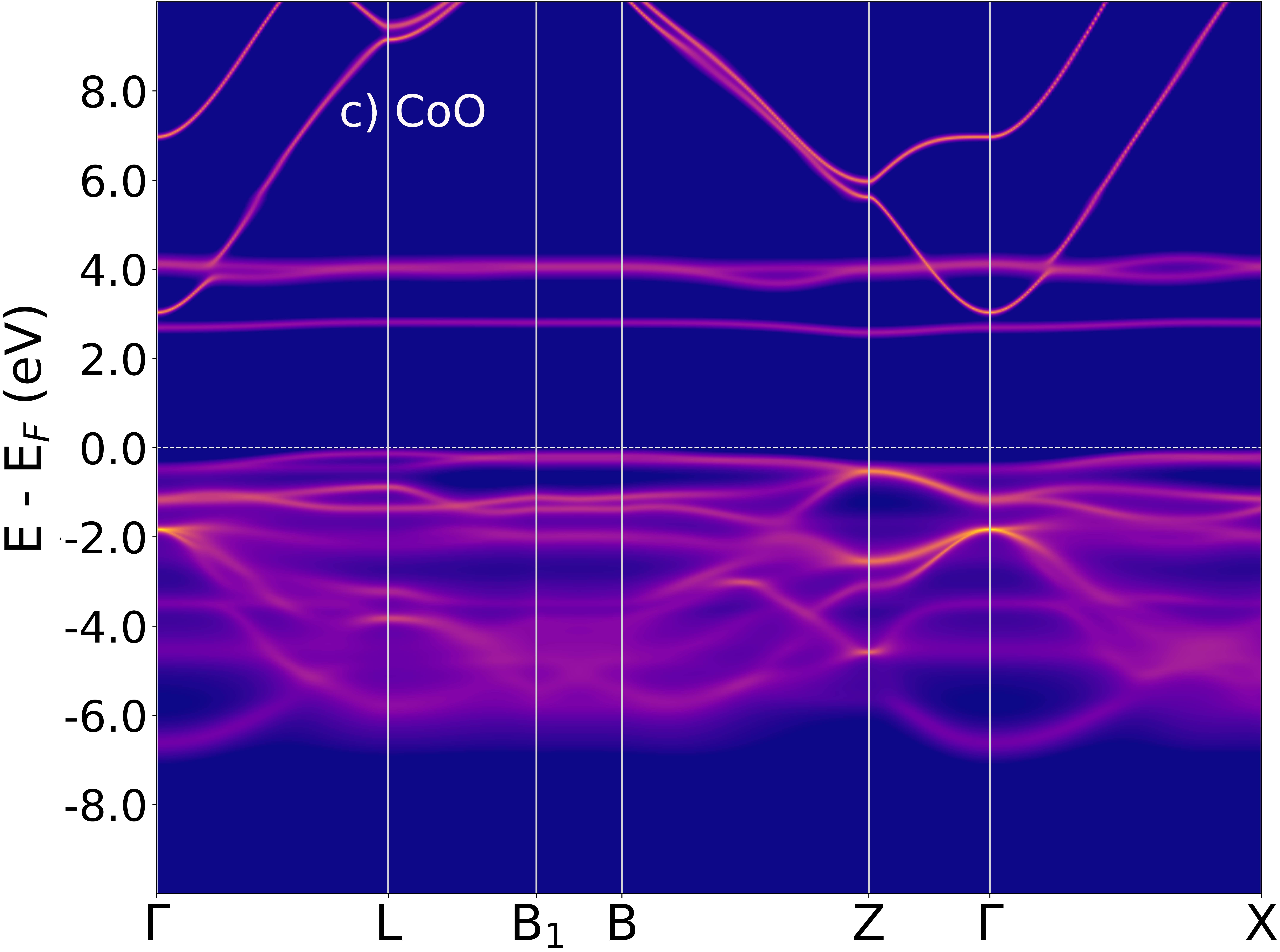}
   \includegraphics[width=0.40\textwidth]{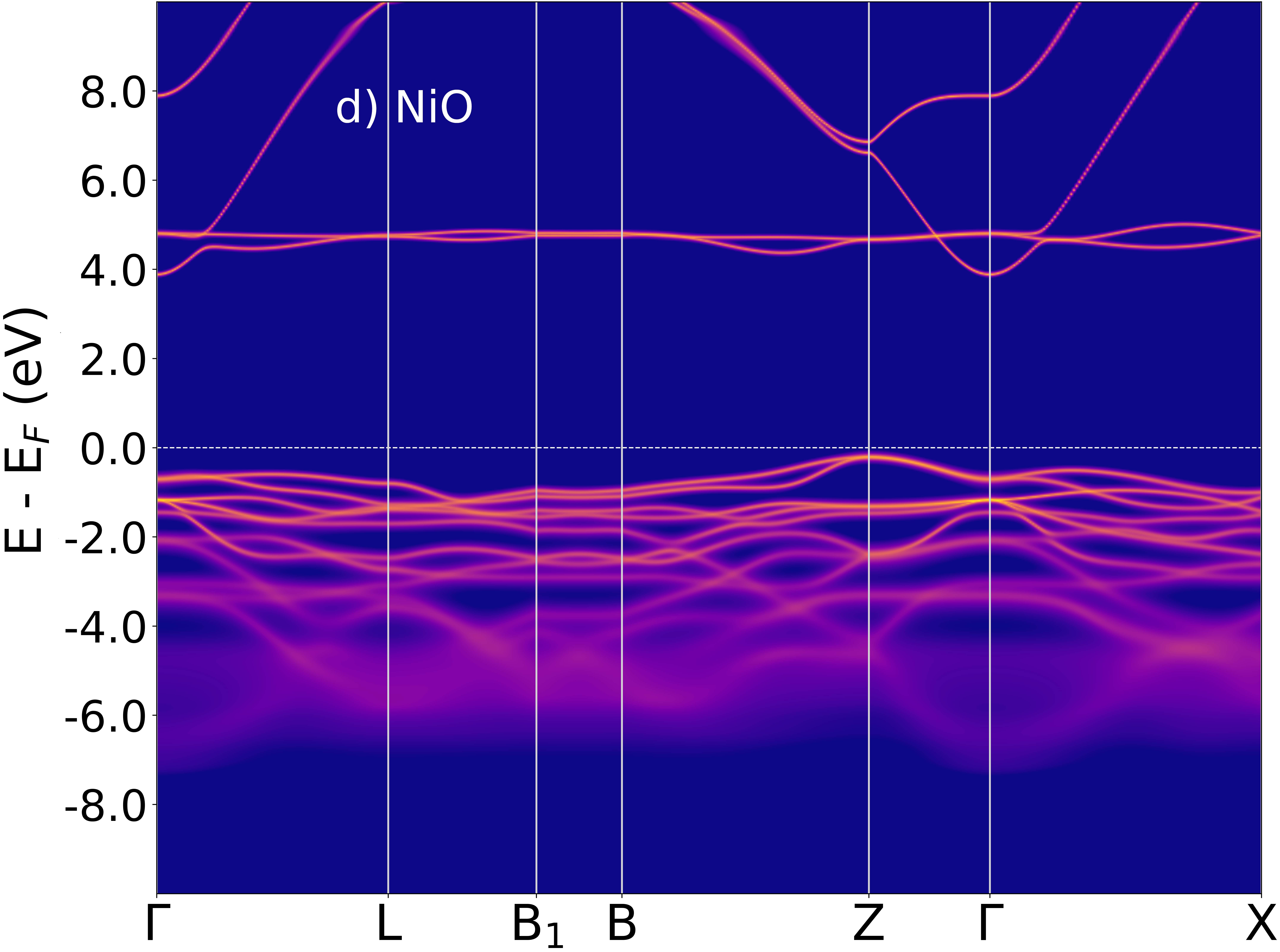}
    \caption{DFT+dynH $k$-resolved spectral functions (colormap) for MnO (a), FeO (b), CoO (c) and NiO (d). The zero of the energy is set at the top of the valence band. A broadening of 0.015 eV has been used.}
    \label{spectral_functions}
\end{figure}
The most relevant difference between the DFT+dynH and DFT+DMFT results~\cite{kunes_nio_2007,mandal_influence_2019}, is the presence of the dispersing band at -8 eV and the AFM flat band at around -4 eV, both close to $\Gamma$. 
They can be labeled as AFM bands as they are missing in non-magnetic DFT calculations, while they appear in the spin-DFT broken symmetry AFM solution~\cite{shen_electronic_1991}. Their vanishing spectral weight in the DFT+DMFT spectrum and absence in the ARPES data of Ref.~\cite{shen_photoemission_1990} were conjectured~\cite{mandal_influence_2019,mandal_systematic_2019} to be a signature of strong correlations.
It is important to note, then, that these states are present both in ARPES data~\cite{kuhlenbeck_molecular_1991,shen_electronic_1991} and in our DFT+dynH results. In fact, in the present work the effective screened interaction is calculated at the first-order in perturbation theory, with a zero-frequency limit of U($0$)= 5.3 eV, which is half of the effective $U$ value used in the DFT+DMFT calculations~\cite{mandal_influence_2019,mandal_systematic_2019}. The slight underestimation of the AFM bands around -4 eV could be a feature of the r2SCAN starting point, a lack of self-consistency or the intrinsically approximate nature of the double-counting term. It is important to note that, when comparing theoretical spectral functions and experimental ARPES points, the lack of signal in the latter can be due to the (vanishing) matrix elements of the dipole operator. Moreover, in the region between -1 eV and -3 eV in Fig.~4 (or Fig.~13) in~\cite{shen_electronic_1991}, other transitions are present, defined as sub-components of the main feature A (see Fig. 11 for their blown-up view). See Page 44 in~\cite{shen_electronic_1991} for further analysis on those components. 

We show in Fig.~\ref{spectral_functions} the $k$-resolved spectral functions for all four monoxides on a standard path. In agreement with DFT+DMFT~\cite{mandal_influence_2019}, it can be seen that MnO possess the sharpest bands, whereas a stronger spectral weight transfer from the QP peaks to the satellites appears in FeO, CoO and NiO.\footnote{Note that this is consistent with the high-spin 5/2 configuration of the d$^5$ shell in Mn$^{2+}$ being more mean-field like.}
We stress the impressive similarity with the spectral functions presented in Ref.~\cite{mandal_influence_2019}, that are obtained with an (expensive) non-perturbative solution of an effective impurity model.

In conclusion, in this work we have 1) generalized the dynamical Hubbard functional of Ref.~\cite{chiarotti_energies_2024} to the treatment of magnetic systems, and 2) applied the framework to study valence and conduction spectral functions of the prototypical Mott-Hubbard/charge-transfer monoxide series of MnO, FeO, CoO and NiO. 
We obtain remarkable agreement with PES/IPES and ARPES experimental data and state-of-the-art DFT+DMFT calculations \cite{mandal_influence_2019,mandal_systematic_2019}, in a transparent and inexpensive formulation.
The fundamental gaps and relative intensities and overall bandwidth of valence and conduction excitations appear to be consistent with experimental data for all four monoxides.
Moreover, the general features of the spectral function for NiO along $\Gamma$-X direction are in close agreement with available experimental data \cite{shen_electronic_1991,kuhlenbeck_molecular_1991}. The dispersing bottom of the valence band and the characteristic flat antiferromagnetic bands are found, at variance with previous calculations~\cite{mandal_influence_2019,kunes_nio_2007}, where these bands show an almost completely vanishing spectral weight. 

Importantly, the present dynamical framework provides access to particle lifetimes, satellites, spectral-weight transfer and bands renormalization, i.e. fundamental signatures of electrons correlations, unavailable to static methods such as hybrid functionals and DFT+U, or to QSG\^W. 
In particular, DFT+dynH, with appealing simplicity and minimal computational cost proves to be an efficient tool for predicting electronic structure properties of complex oxides with various degrees of symmetry-broken orbital orderings and hybridizations, thus posing itself as a promising tool for simulating realistic correlated materials, that may be prohibitive for methods employing exact solutions.

This work has been possible thanks to a Bosch Stiftung grant.
T.C. and N.M. also acknowledge the support from the Swiss National Science Foundation (SNSF) through Grant No. 200020\_213082 for this work.
The authors wish to thank Lucia Reining, Max Amsler, Georgy Samsonidze and Thomas Eckl for many fruitful discussions.

\renewcommand{\emph}{\textit}
\bibliographystyle{apsrev4-1}
\bibliography{references,references_extra}

\pagebreak

\setcounter{equation}{0}
\setcounter{figure}{0}
\setcounter{table}{0}
\setcounter{section}{0}
\setcounter{page}{1}
\makeatletter
\renewcommand{\theequation}{S\arabic{equation}}
\renewcommand{\thefigure}{S\arabic{figure}}
\renewcommand{\thetable}{S\arabic{table}}
\renewcommand{\thesection}{S\arabic{section}}

\onecolumngrid\clearpage

\title{Supplemental Material: \\
        Dynamical Hubbard approach to correlated materials:\\
    the case of transition-metal monoxides}
\maketitle

\section{DFT electronic structure}
\begin{figure*}[!ht]
    \includegraphics[width=0.35\textwidth]{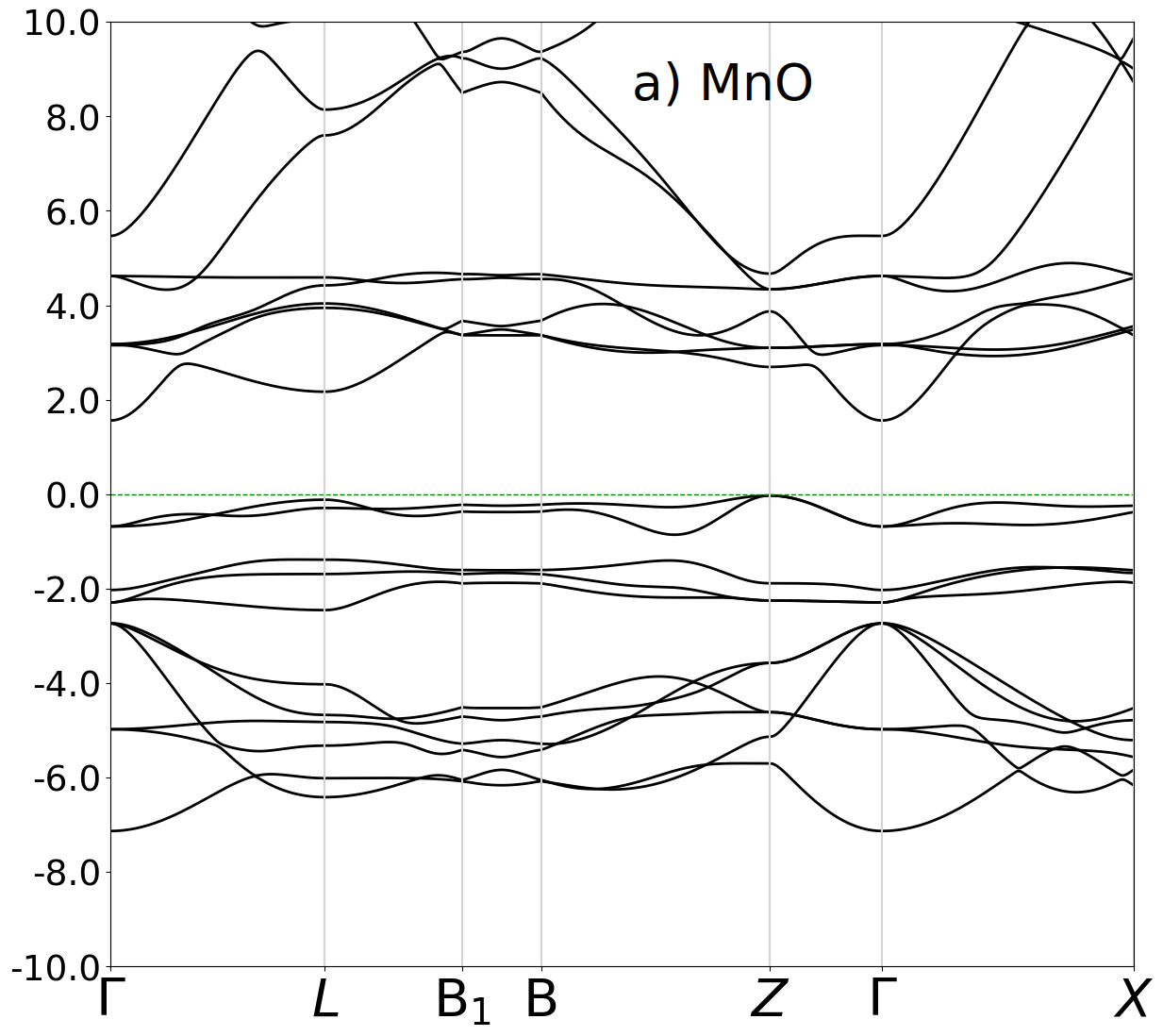}   \includegraphics[width=0.35\textwidth]{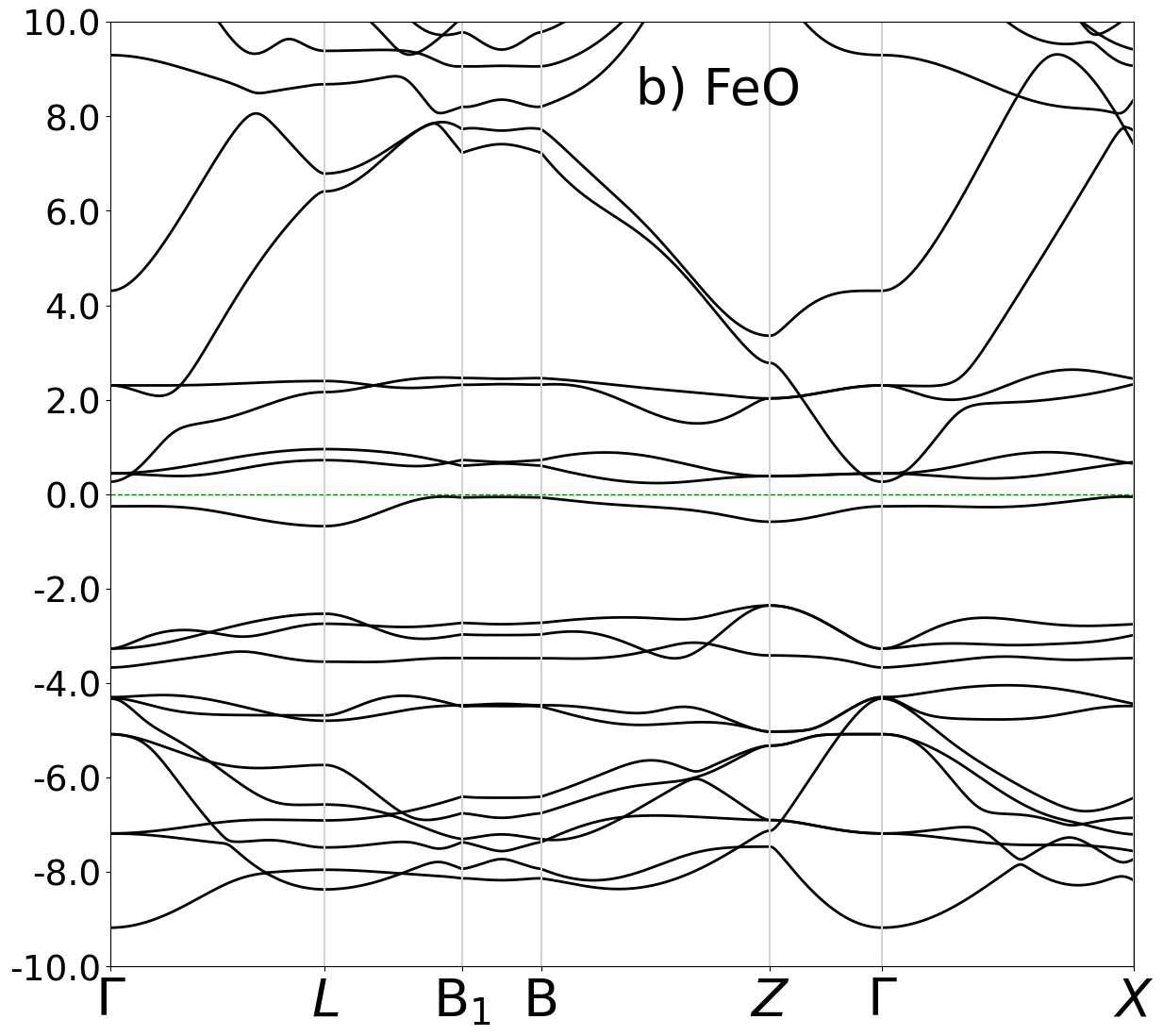}
    \includegraphics[width=0.35\textwidth]{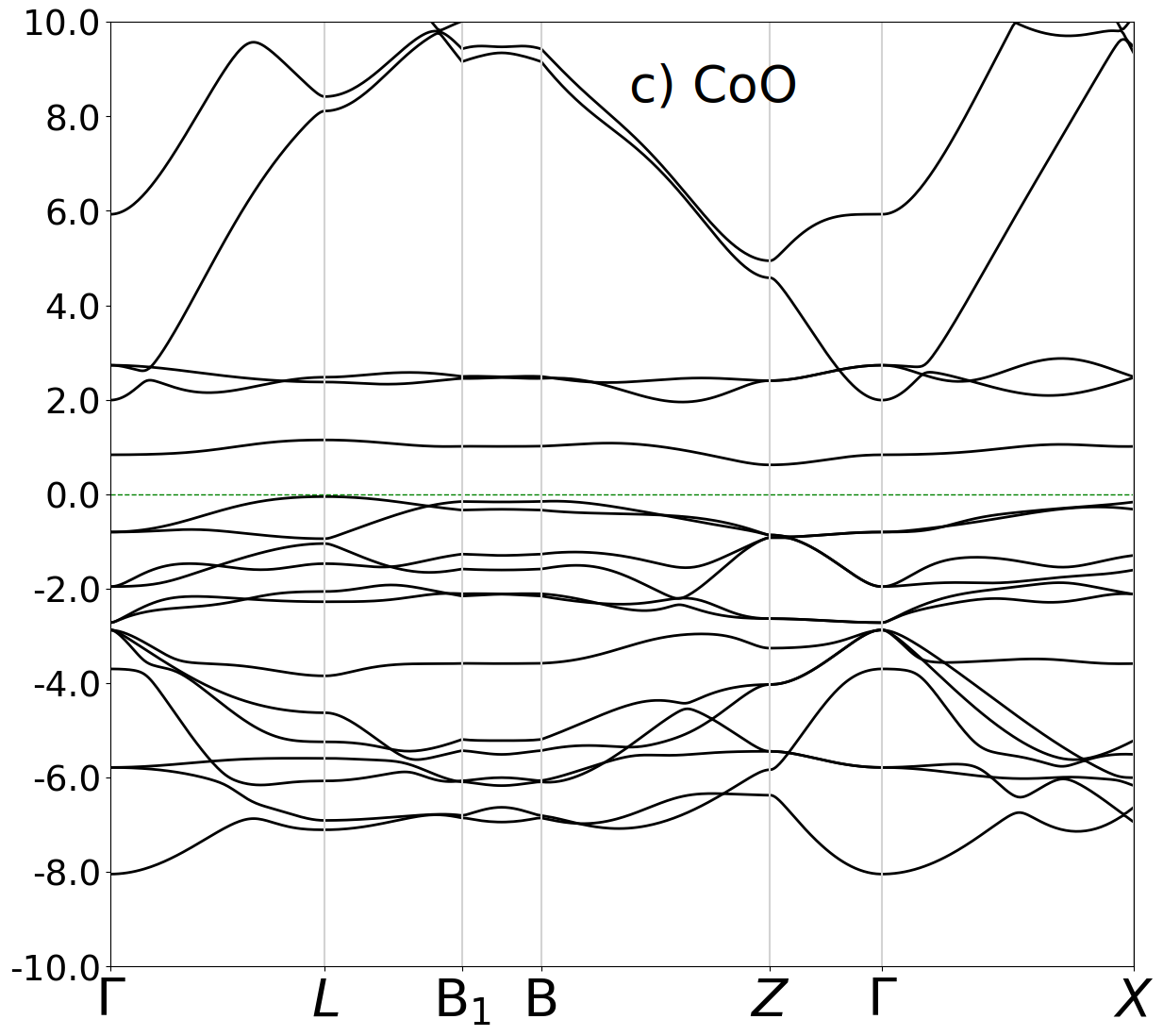}
    \includegraphics[width=0.35\textwidth]{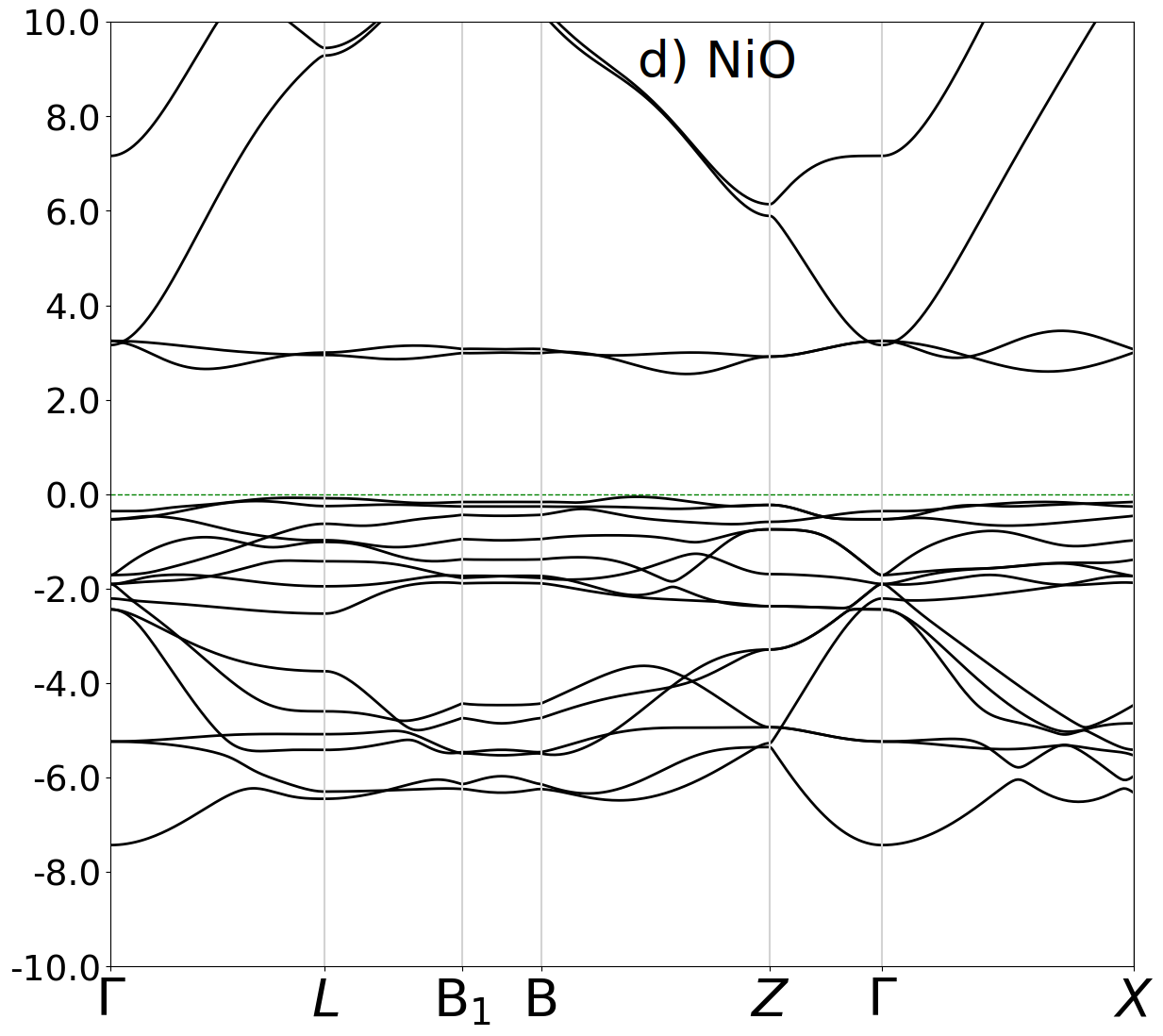}

    \caption{\small r2SCAN DFT band structure for MnO (a), FeO (b), CoO (c) and NiO (d). The Fermi energy is set at top of the valence band.}
    \label{band_structure}
\end{figure*}

\begin{figure*}[!ht]    
    \includegraphics[width=0.49\textwidth]{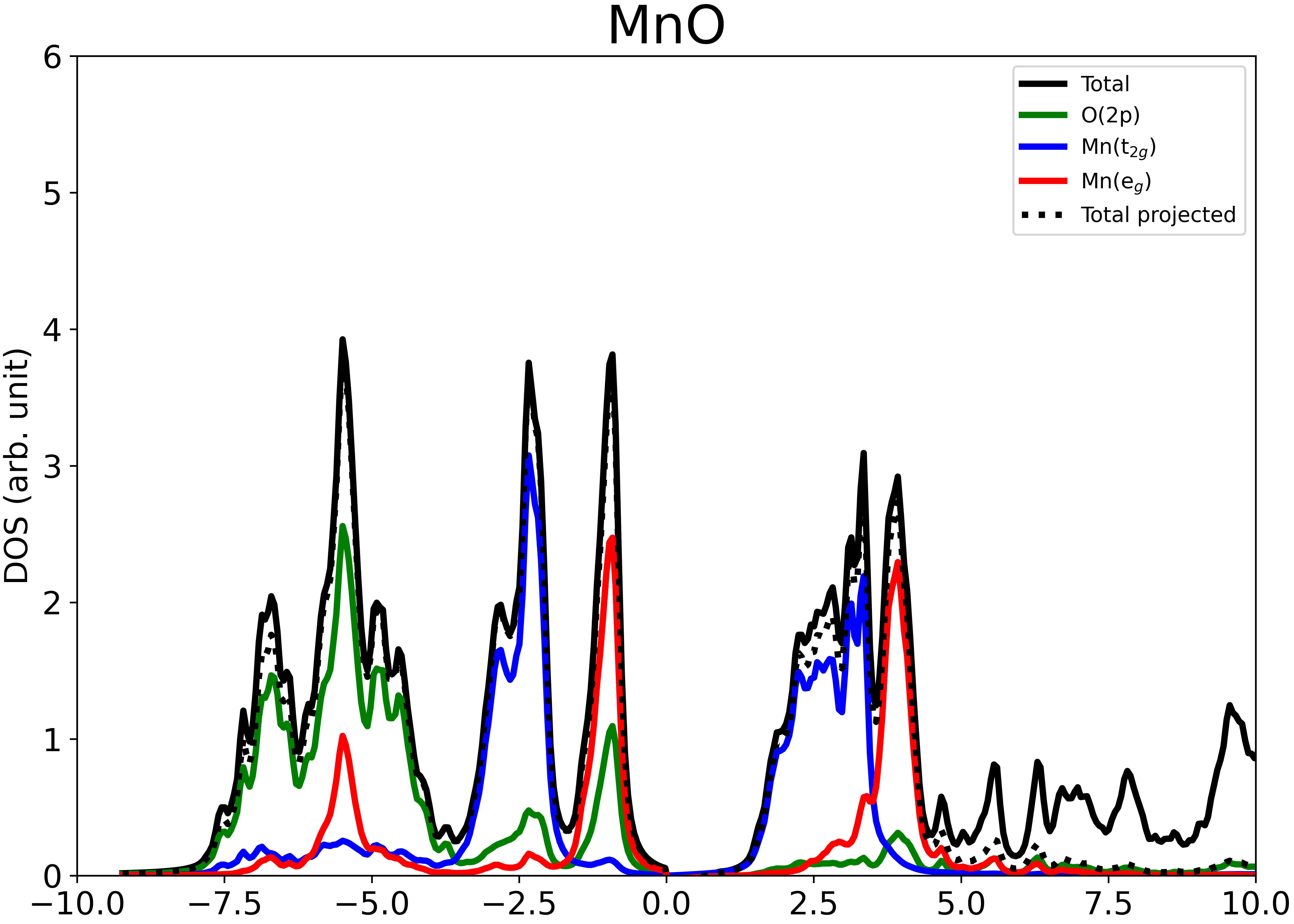}   \includegraphics[width=0.49\textwidth]{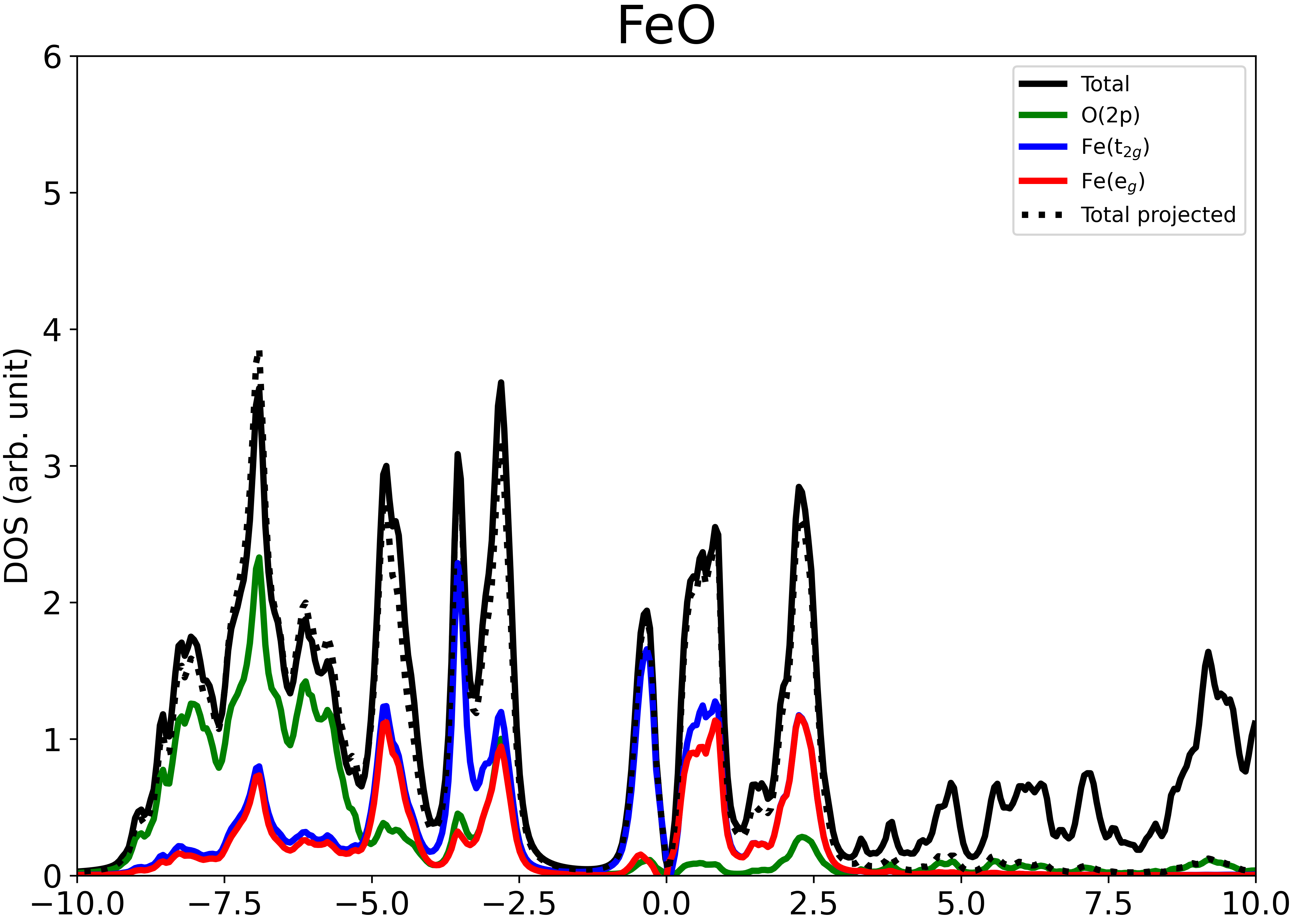}
    \includegraphics[width=0.49\textwidth]{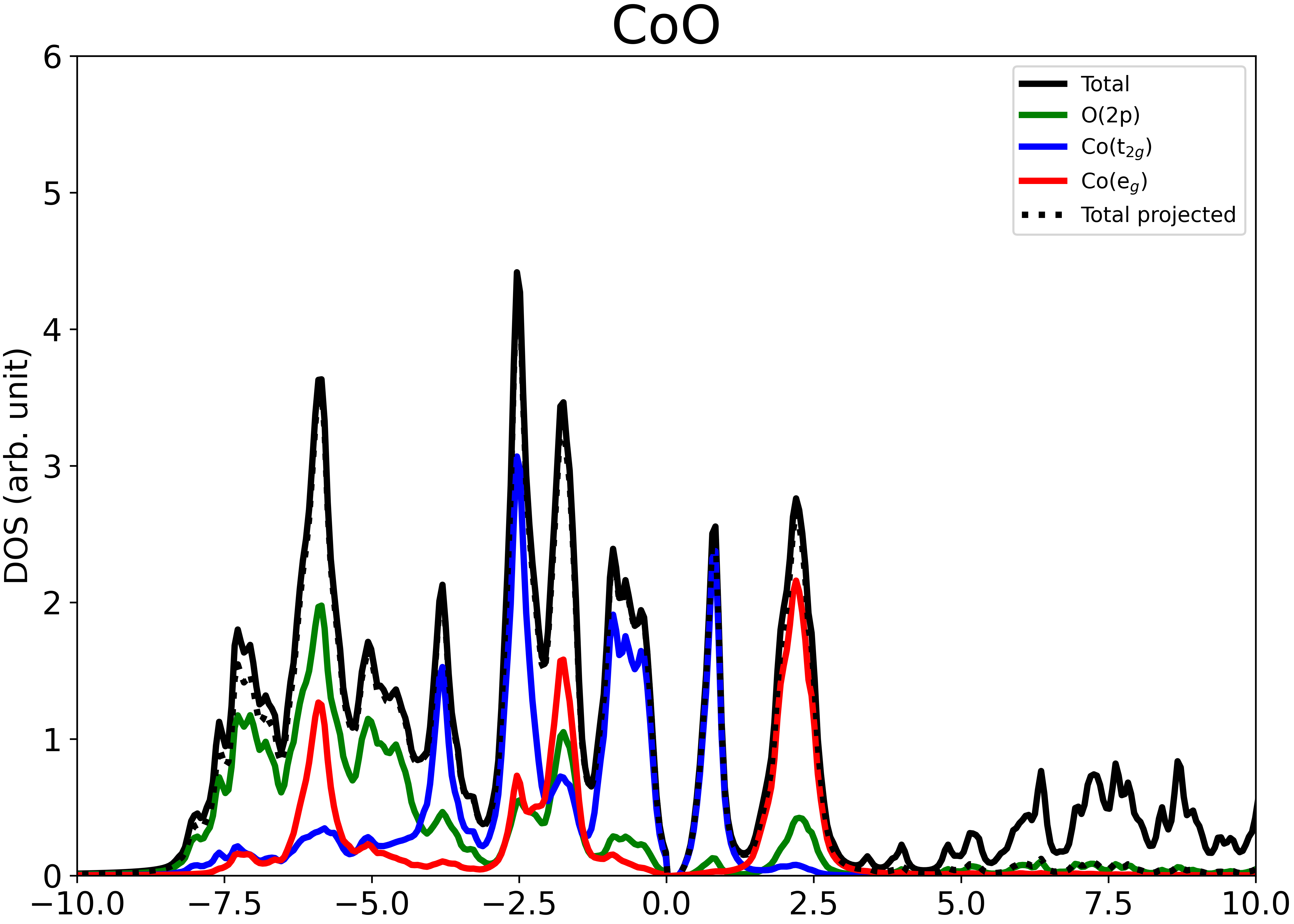}    
    \includegraphics[width=0.49\textwidth]{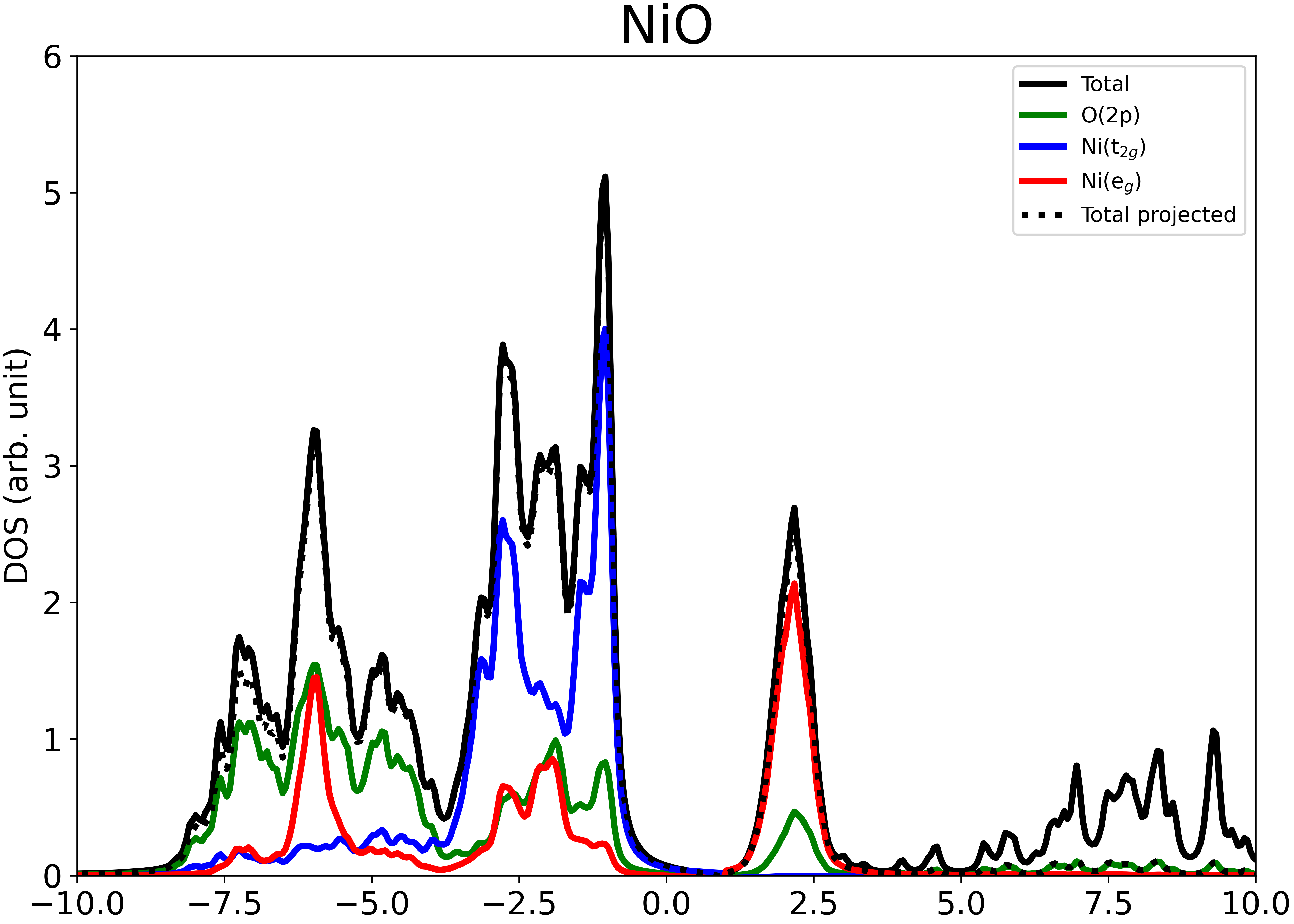}
    \caption{\small Projected spin density of states for r2SCAN for MnO (a), FeO (b), CoO (c) and NiO (d). A broadening of 0.05 eV has used.}
    \label{proj_dosses_r2s}
\end{figure*}

In Fig.~\ref{band_structure} we show the DFT band structures for MnO (a), FeO (b), CoO (c) and NiO (d). The considered crystal structure is the rhombohedral
(R3m) with the AFM-II magnetic ordering
along [111] direction, with two transition metal ions per the unit
cell. The respective lattice constants were set as 4.445 {\AA}~\cite{johnston1956study}, 4.334 {\AA}~\cite{mccammon1984effects}, 4.254 {\AA}~\cite{carey1991preparation}, and 4.171 {\AA}~\cite{bartel1971exchange}. 
With the r2SCAN exchange-correlation functional~\cite{furness_accurate_2020} all four monoxides are insulating in DFT. 
The calculated local magnetic moments are 4.6, 3.7, 2.7 and 1.78 $\mu$B per transition metal atom, which compare well with the experimental data of 4.58~\cite{cheetham1983magnetic},  4.0~\cite{fjellvaag1996crystallographic}, 3.8~\cite{herrmann1978equivalent}, 1.9~\cite{cheetham1983magnetic} $\mu$B, respectively, with CoO the only one underestimated by more than 1.0 $\mu$B \footnote{Albeit 2.7 $\mu$B is more consistent with the Co$^{2+}$ high-spin configuration, which would imply 3 $\mu$B}.
In Figure~\ref{proj_dosses_r2s} the r2SCAN projected density of states are shown. It can be seen that MnO, CoO and NiO all share already a certain degree of hybridization between TM $3d$ states and oxygen $2p$ states, at the top of the valence band, FeO being the only one of the four showing a $d-d$ kind of electronic gap. The effect of the DFT+dynH correction (Fig.~\ref{proj_dosses}) is to increase that hybridization in the valence band, again with the exception of FeO, while the oxygens' weight remains negligible in the conduction states in all four compounds. The degree of the Mott-Hubbard/charge-transfer excitation in the series is then given by the different redistribution of the atomic characters in the valence band only, going from a considerable charge-transfer nature of MnO and NiO especially, to a lower degree in CoO and a pure Mott-Hubbard for the FeO gap. 

\section{Sum-over-poles representation of the local screened Coulomb interactions $U(\omega)$}
From the r2SCAN ground-state obtained (see previous Section), the RPA polarization and local screened Coulomb interactions are calculated.
\begin{figure*}[!ht]
    \includegraphics[width=0.42\textwidth]{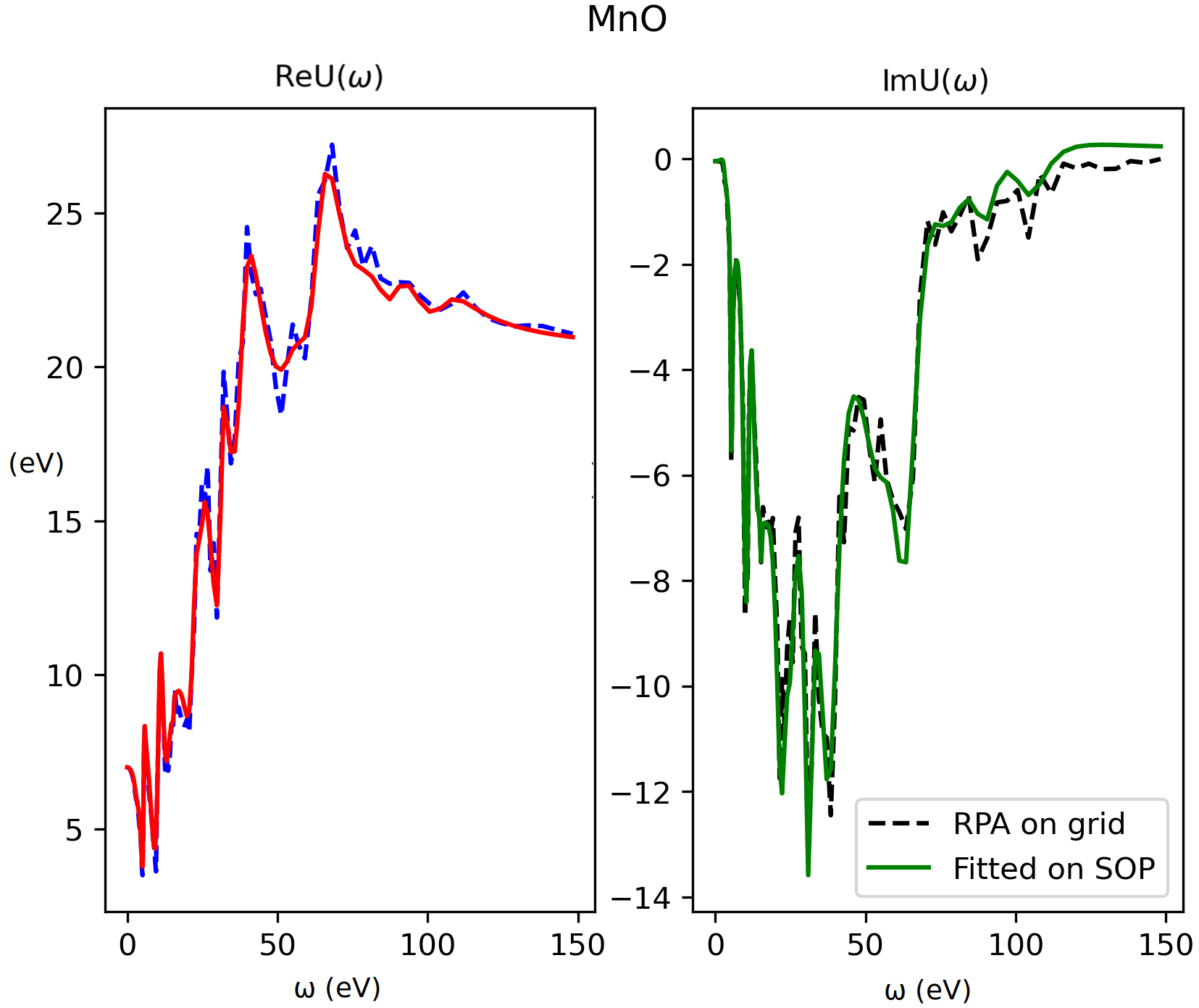}   \includegraphics[width=0.42\textwidth]{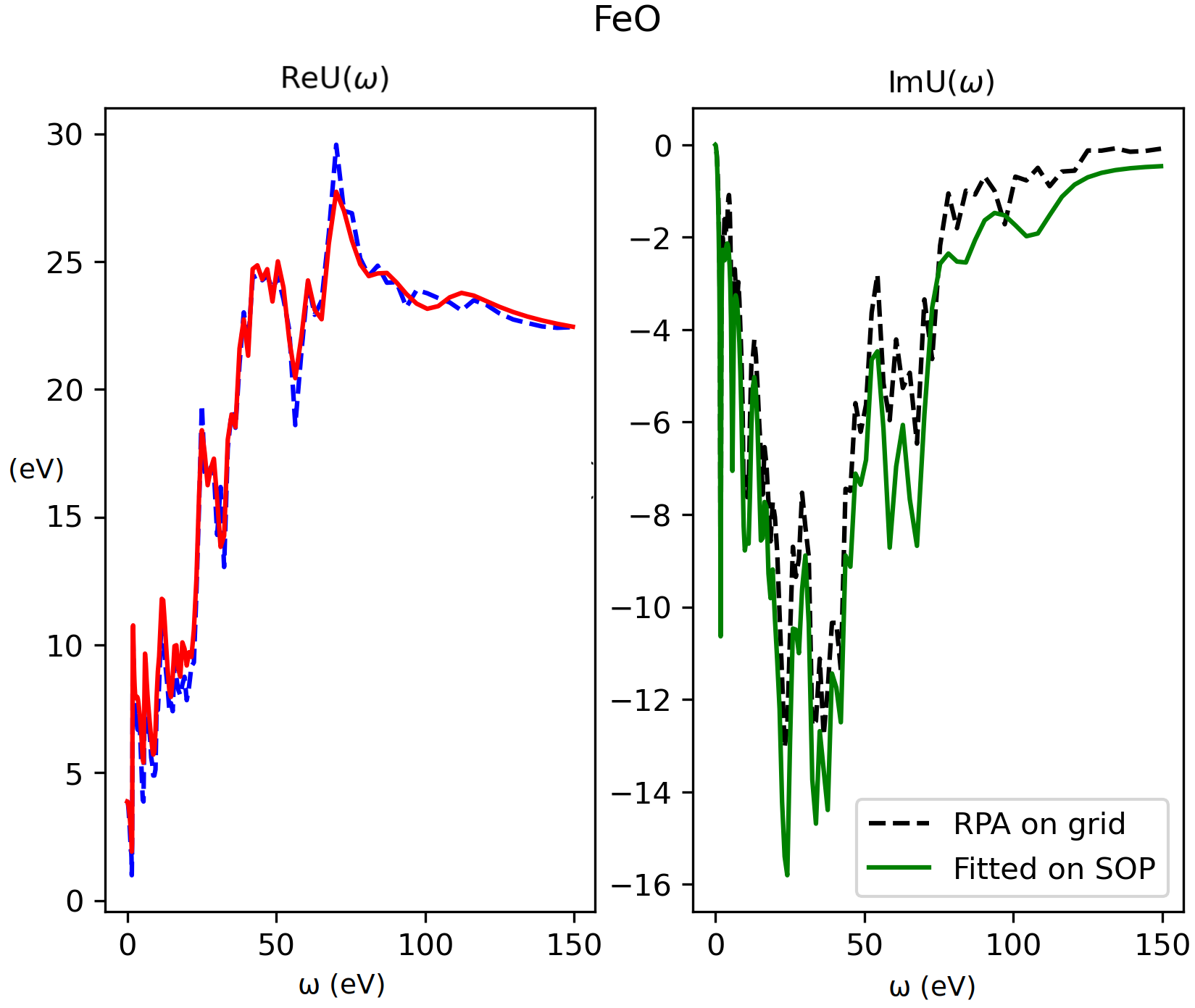}
    \includegraphics[width=0.42\textwidth]{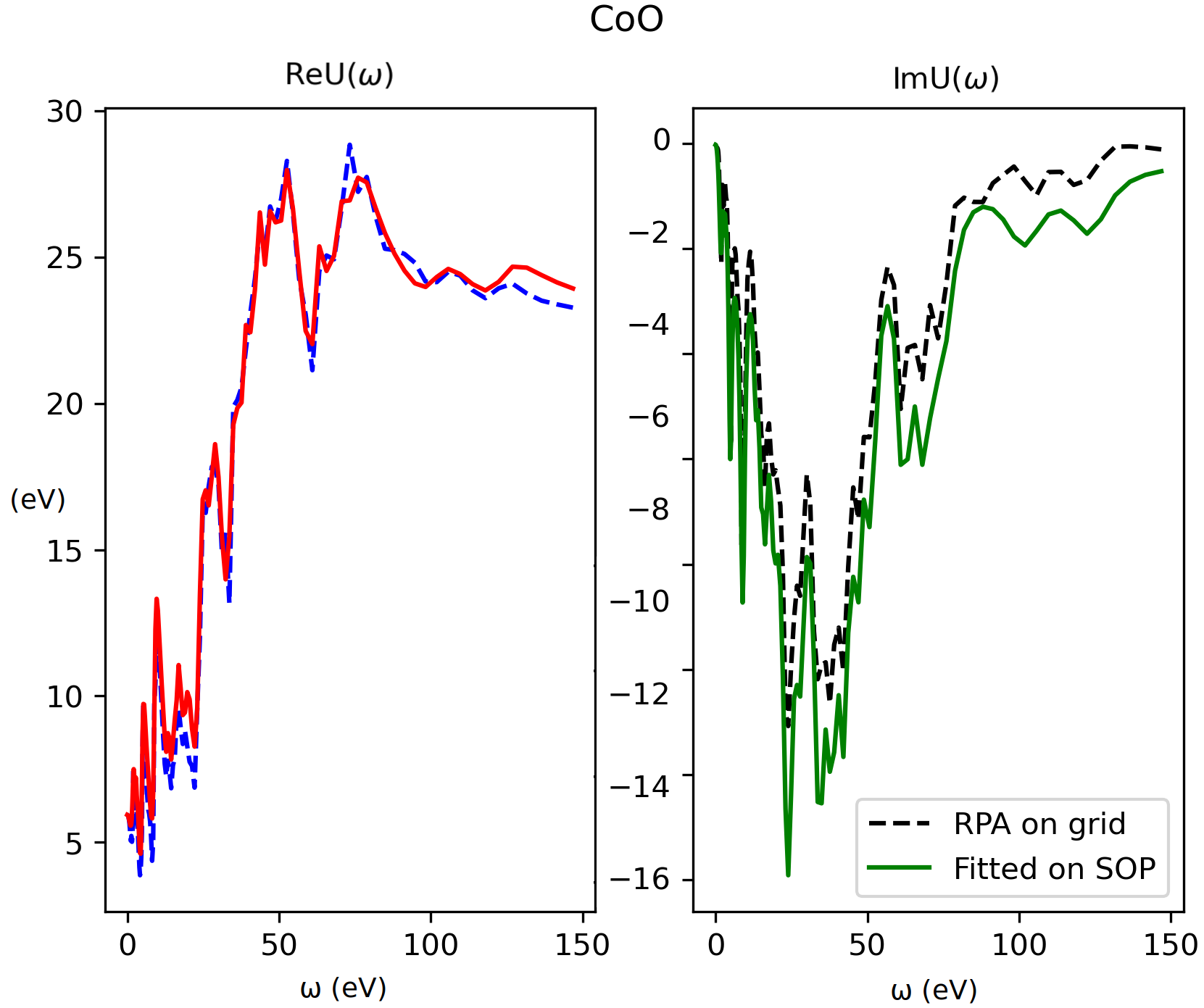}
    \includegraphics[width=0.42\textwidth]{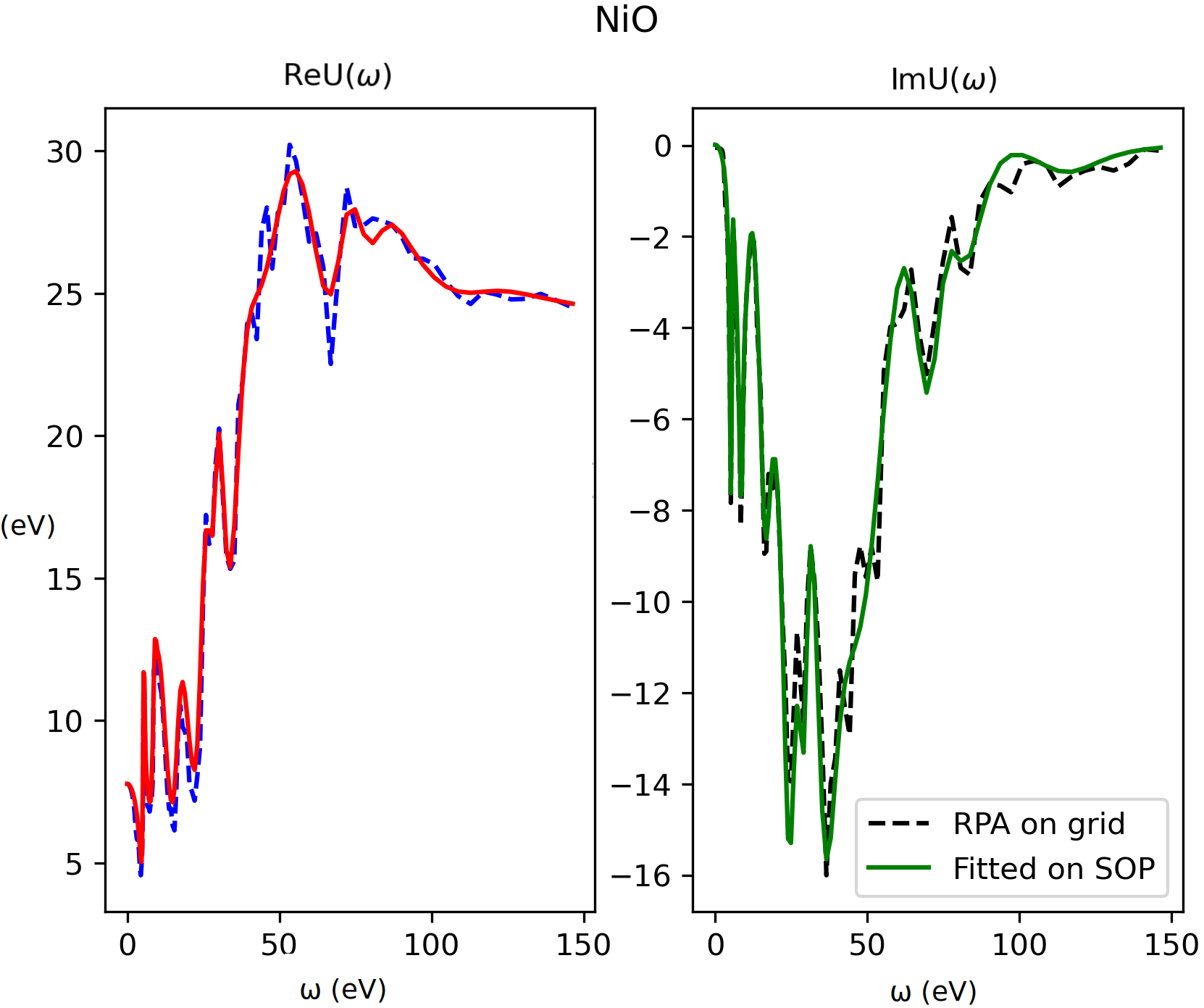}

    \caption{\small $U(\omega)$ and SOP fitting for MnO, FeO, CoO and NiO.}
    \label{u_omega_fits}
\end{figure*}
To facilitate the calculation of the average on-site interaction,
\begin{multline}
    U_I(\omega) = \frac{1}{N_d} \sum_{\substack{mm' \\ \sigma \sigma'}} \\ \bra{\phi_{m,I}^{\sigma} \phi_{m',I}^{\sigma'}} W_{RPA}(\mathbf{r},\mathbf{r'},\omega)\ket{\phi_{m,I}^{\sigma}\phi_{m',I}^{\sigma'}},
     \label{eq:U_average}
\end{multline}
we assume that the sum of
the two parallel-spin elements on the two transition-metal sites of the unit cell is equal to that of the two anti-parallel-spin elements at the same sites.
Explicitly, we have:
\begin{multline}
    \bra{\phi_{m,I_1}^{\uparrow} \phi_{m',I_1}^{\uparrow}} W\ket{\phi_{m,I_1}^{\downarrow}\phi_{m',I_1}^{\downarrow}} + \\  \bra{\phi_{m,I_1}^{\downarrow} \phi_{m',I_1}^{\downarrow}} W\ket{\phi_{m,I_1}^{\uparrow}\phi_{m',I_1}^{\uparrow}} \approx \\
    \bra{\phi_{m,I_1}^{\uparrow} \phi_{m',I_1}^{\uparrow}} W \ket{\phi_{m,I_1}^{\uparrow}\phi_{m',I_1}^{\uparrow}} +  \\
    \bra{\phi_{m,I_2}^{\uparrow} \phi_{m',I_2}^{\uparrow}} W\ket{\phi_{m,I_2}^{\uparrow}\phi_{m',I_2}^{\uparrow}}  
    \label{afm_approx}
\end{multline} 

This approximation is justified by the similarity in shape between the spin-up $\ket{\phi_{m,I_1}^{\uparrow}} $
and spin-down $\ket{\phi_{m,I_1}^{\downarrow}}$ Wannier functions on the same site and the identity of the spin-up $\ket{\phi_{m,I_2}^{\uparrow}} $ on one TM site $I_1$
and spin-down $\ket{\phi_{m,I_2}^{\downarrow}}$ on the other TM site $I_2$, given by the antiferromagnetic order of the monoxides ground state. 
By exploiting Eq.~\eqref{afm_approx} for the calculation of the local screened interaction of Eq.~\eqref{eq:U_average} one can limit the matrix elements of $W$ from only one spin channel, averaging them over the two sites.
It can be shown that the difference between the lhs and rhs of Eq.~\eqref{afm_approx} is of the order $(\varepsilon+\varepsilon^2)^2$, with $\varepsilon$ a small parameter capturing the difference in spread of the Wannier functions $\ket{\phi_{m,I_1}^{\uparrow}} $ and $\ket{\phi_{m,I_1}^{\downarrow}}$, and thus it is a good approximation for spin-dependent MLWF matrix elements in these antiferromagnetic systems.
As a computational note, in this work we fit
the local screened Coulomb interaction $U(\omega)$  (see Fig.~\ref{u_omega_fits}) to a sum over poles---as in Ref.~\cite{chiarotti_energies_2024}.
Specifically, we minimize a least-squares cost function from the grid representation and the SOP representation.
This SOP for $U(\omega)$ is then fed to the dynamical Hubbard code---here extended to treat the multi-site magnetic case as explained in the main text.
\begin{figure*}[!ht]
    \includegraphics[width=0.35\textwidth]{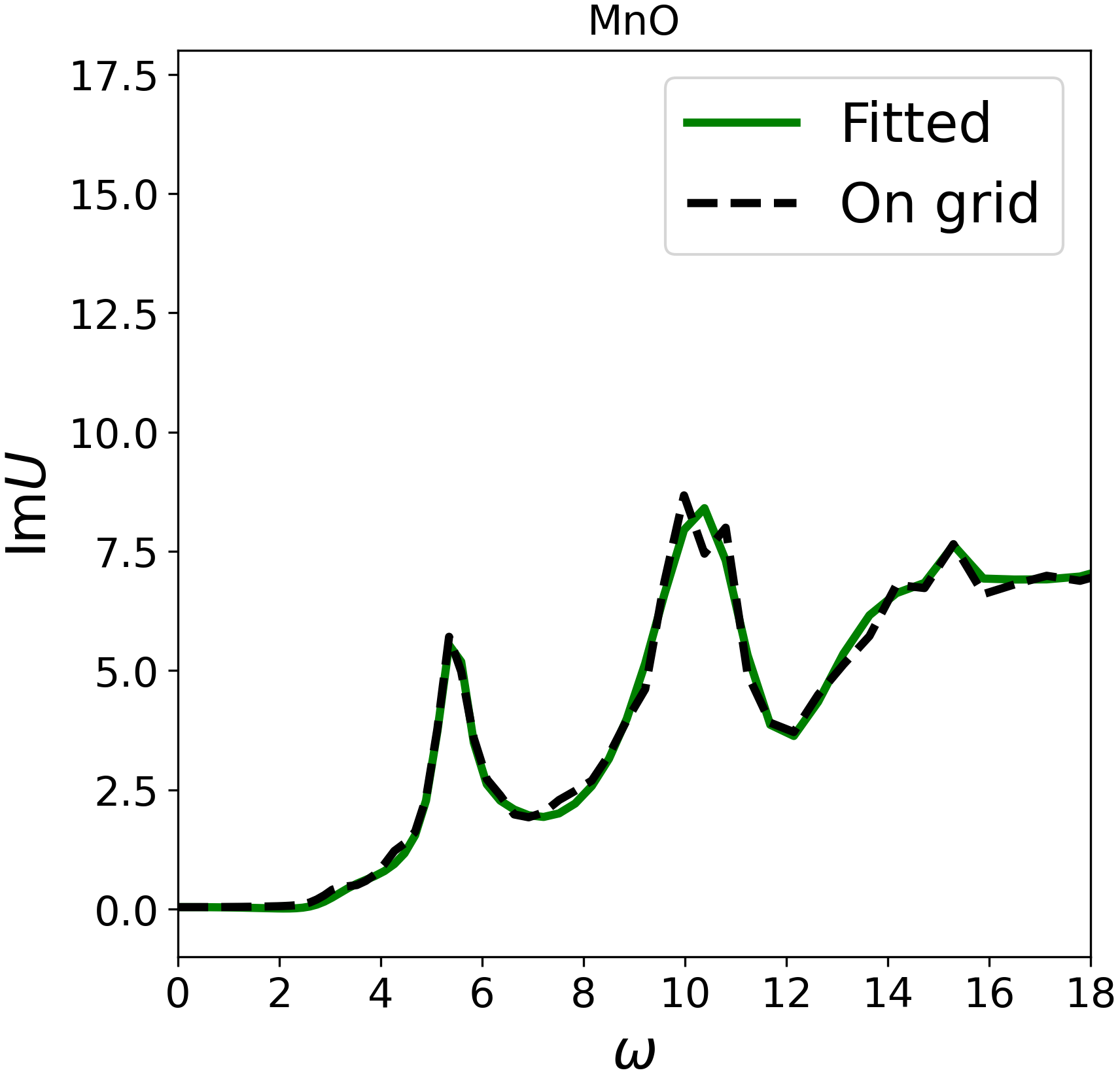} 
    \includegraphics[width=0.35\textwidth]{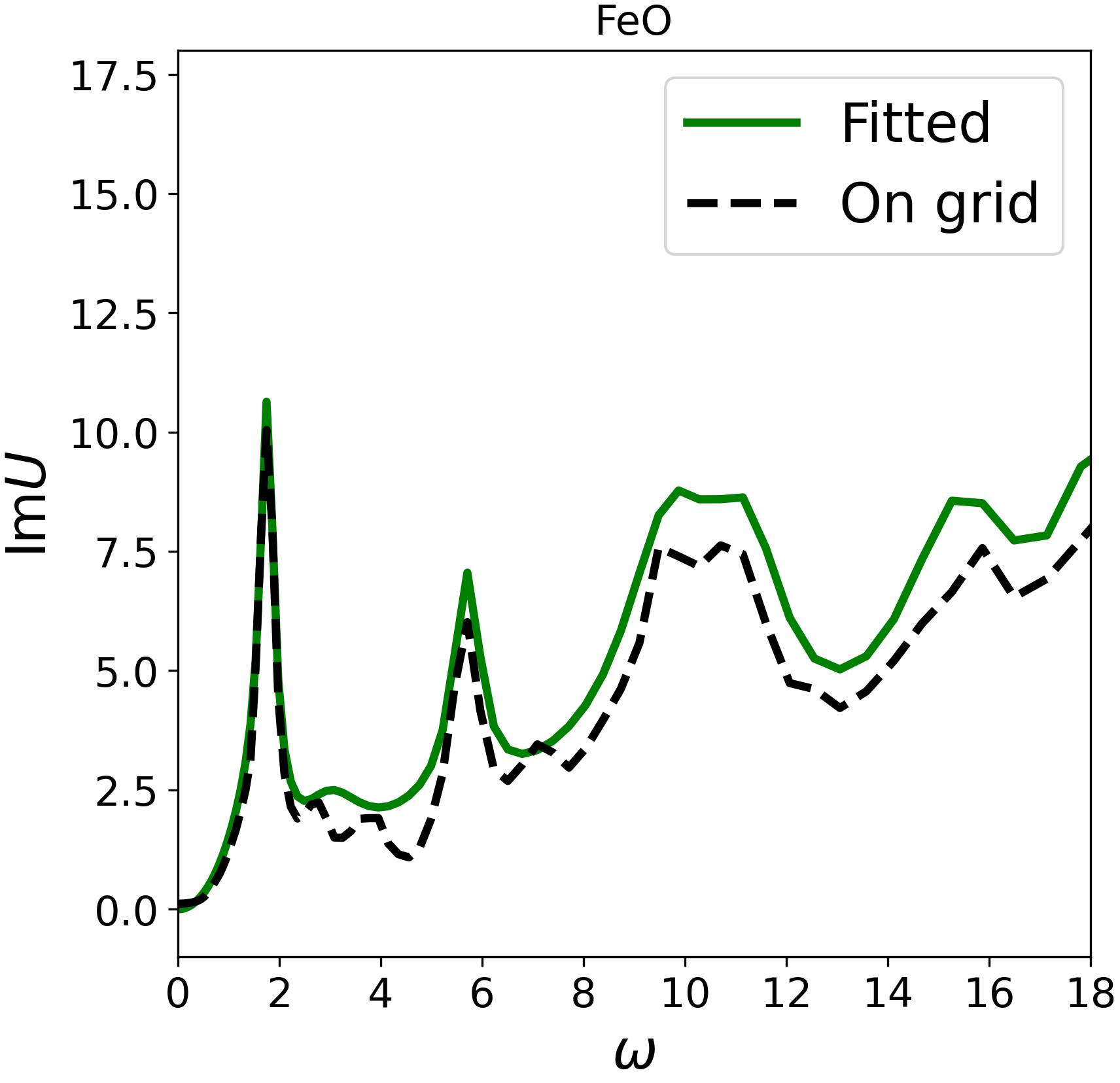}    
    \includegraphics[width=0.35\textwidth]{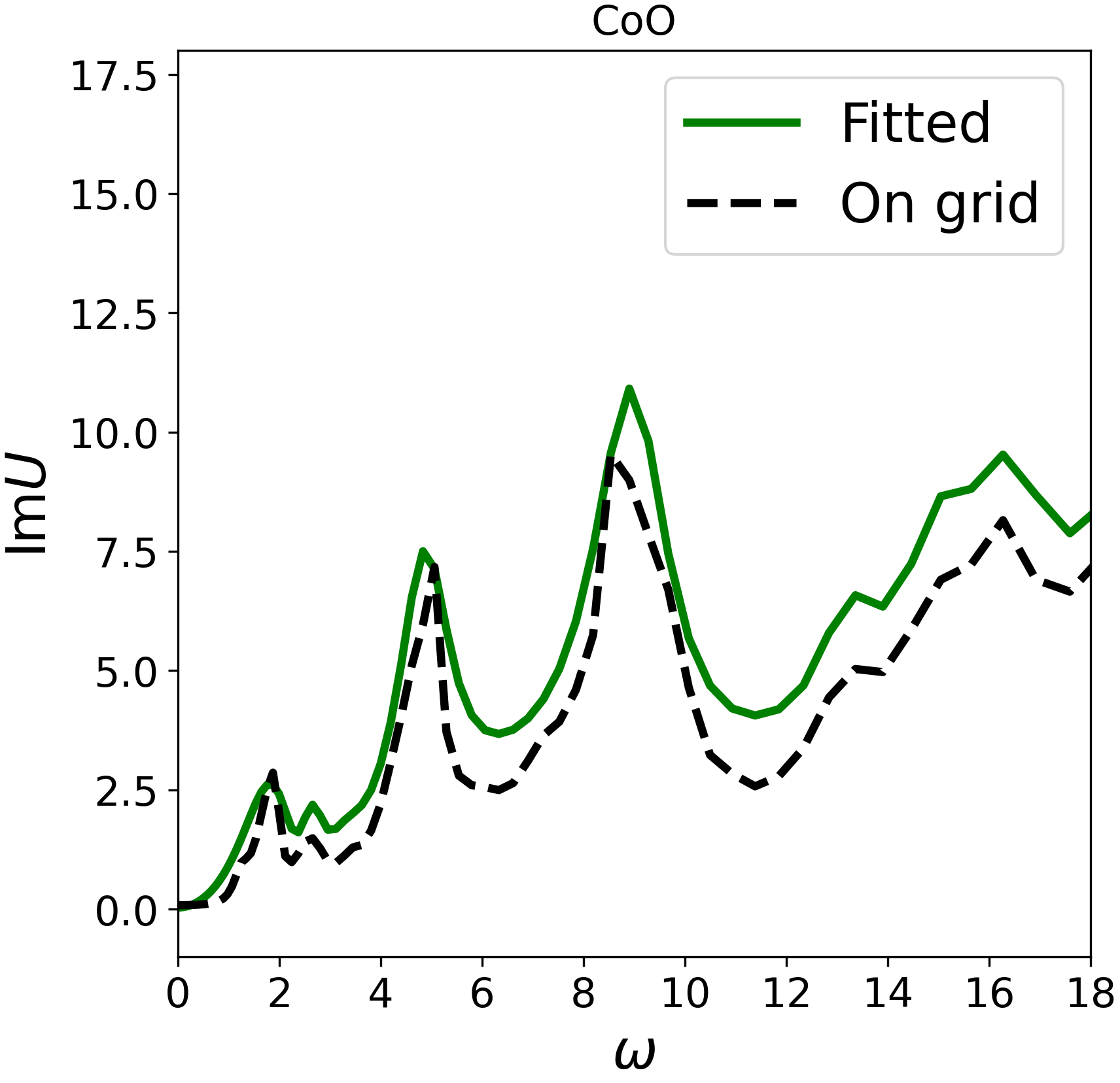}
    \includegraphics[width=0.35\textwidth]{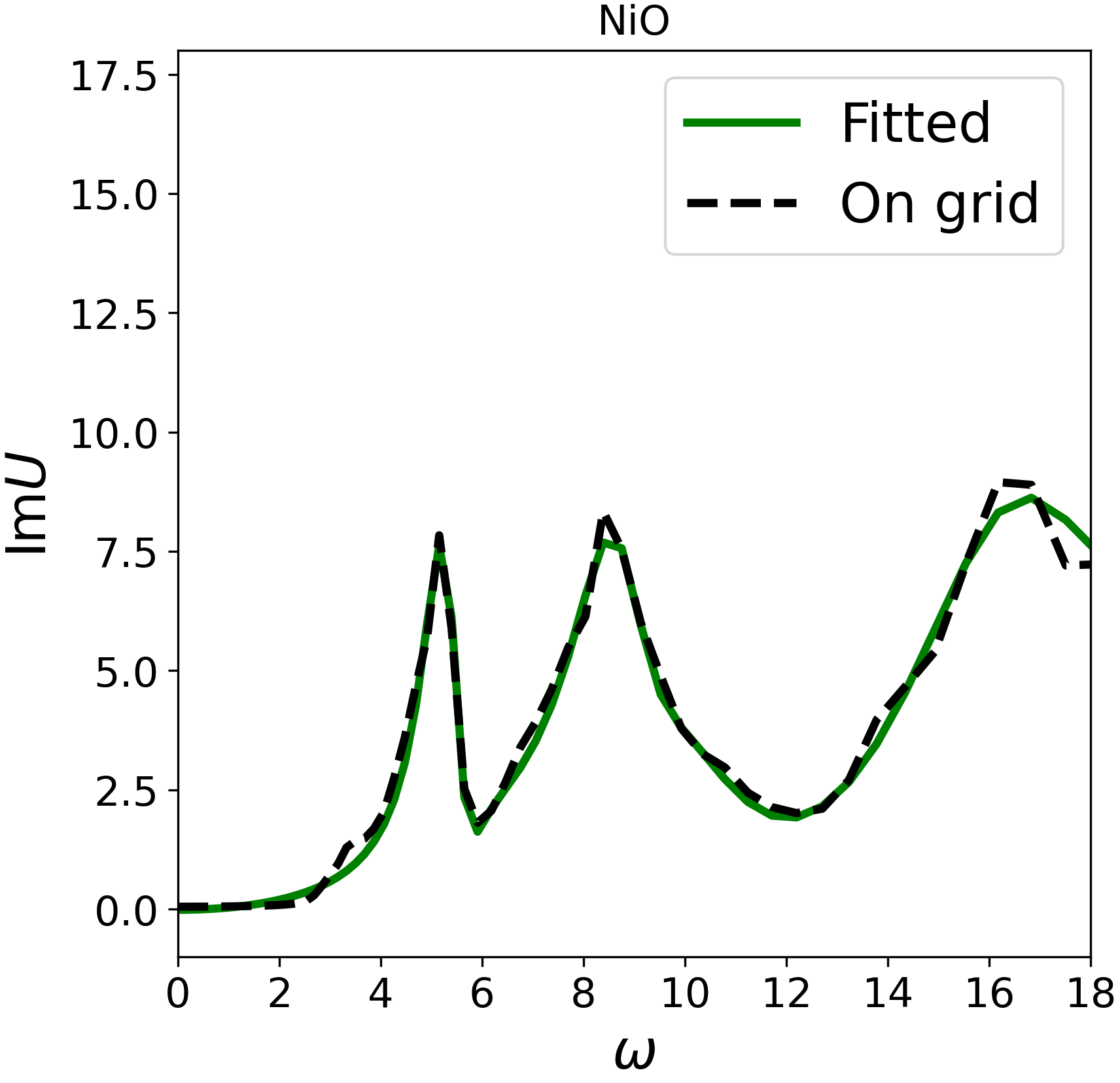}

    \caption{\small Low frequency range of the imaginary part of $U(\omega)$ and SOP fitting for MnO, FeO, CoO and NiO.}
    \label{u_omega_fits_low_frequency_imag}
\end{figure*}
In Fig.~\ref{u_omega_fits_low_frequency_real} and Fig.~\ref{u_omega_fits_low_frequency_imag} the low-energy part of the screened interactions can be seen. The zero-frequency limit of the real part, is the value responsible for the entity of the scissor effect involved in the opening of gaps. The values for the four compounds are respectively 5.97 eV (MnO), 4.8 eV (FeO), 5.4 eV (CoO), 7. 77eV (NiO). They are related to the value of the gap of the starting ground-state, as a bigger gap leads to smaller screening from the RPA response function. The presence of non-vanishing but not too big imaginary parts at low-energies, is related instead to the intensity of the spectral weight transfer from the quasi-particle peaks to the satellites. It can be seen than the first structures in FeO, CoO and NiO appear at smaller energies with respect to the MnO ones, which is consistent with the results of a reduced spectral weight transfer in the latter compound.

\begin{figure*}[!ht]
  \includegraphics[width=0.35\textwidth]{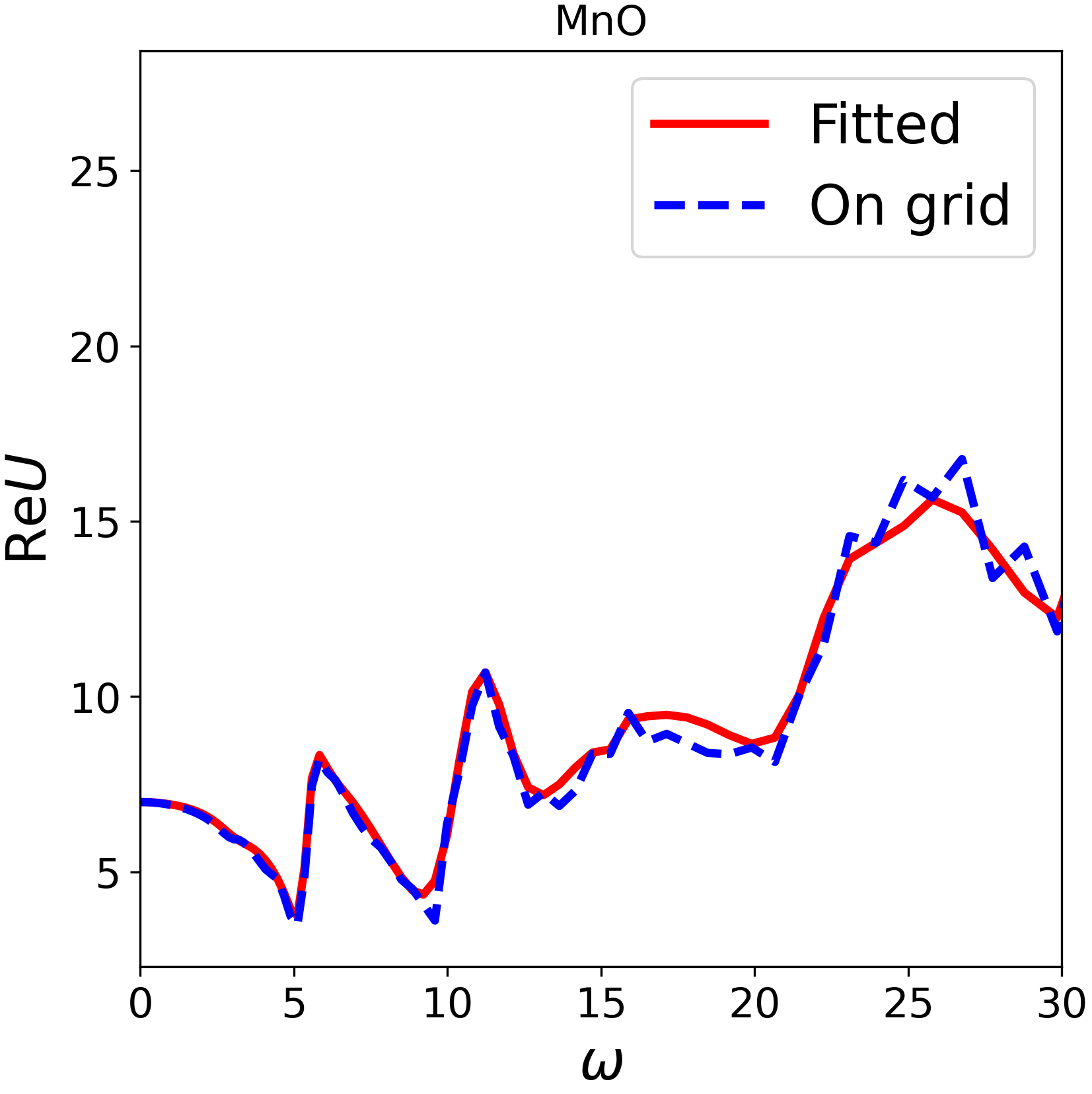}
     \includegraphics[width=0.35\textwidth]{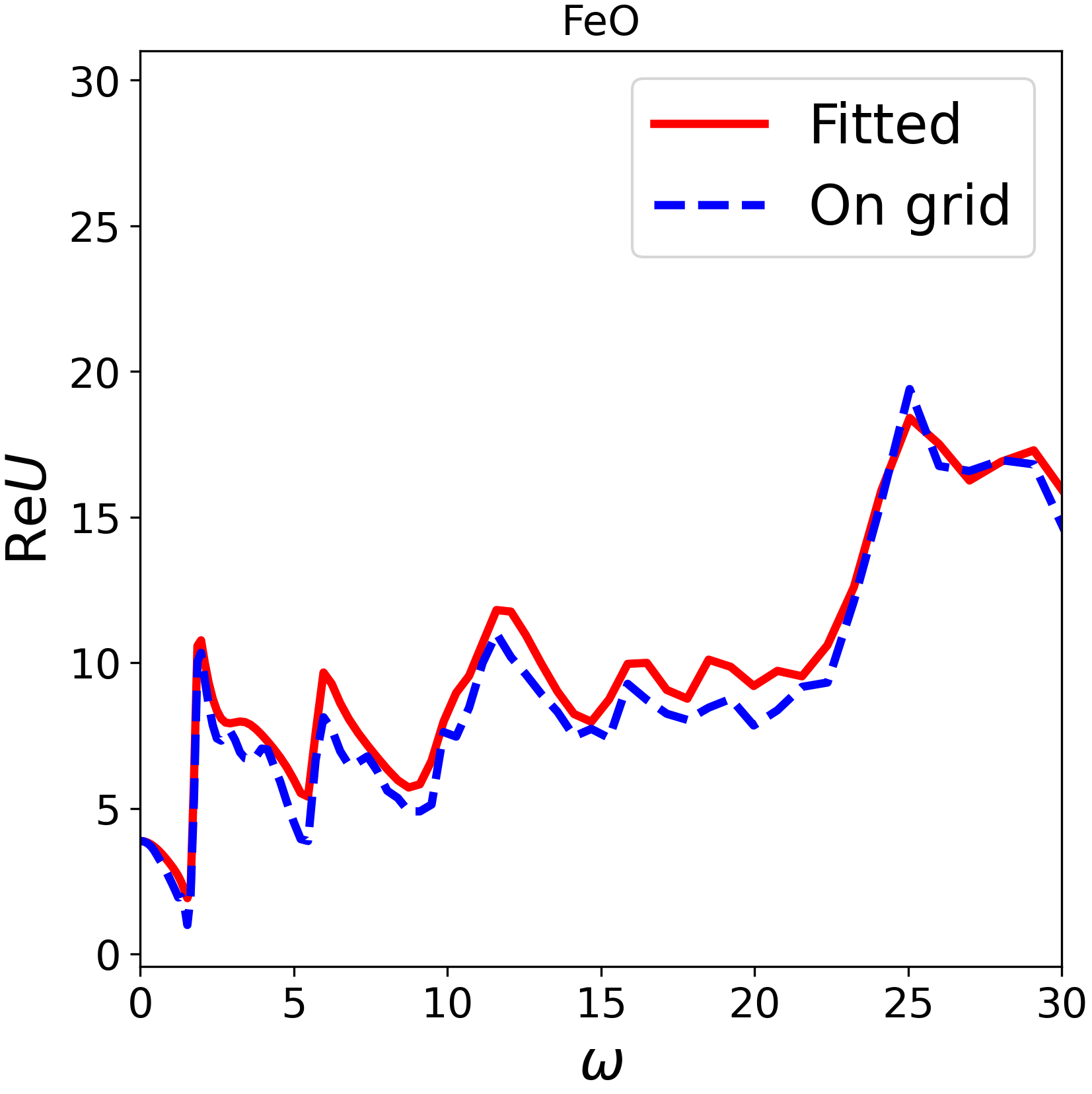}
         \includegraphics[width=0.35\textwidth]{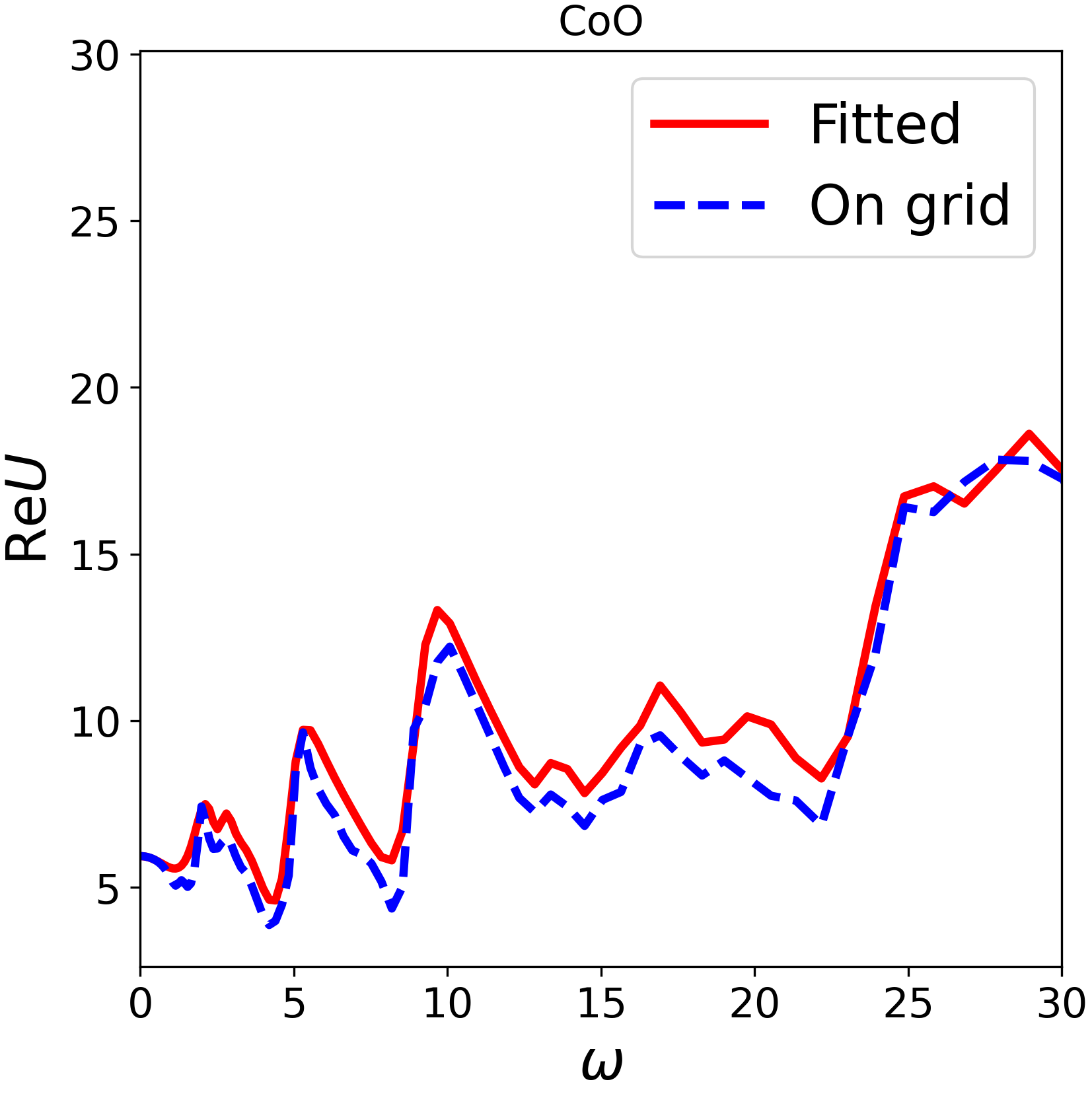}
    \includegraphics[width=0.35\textwidth]
    {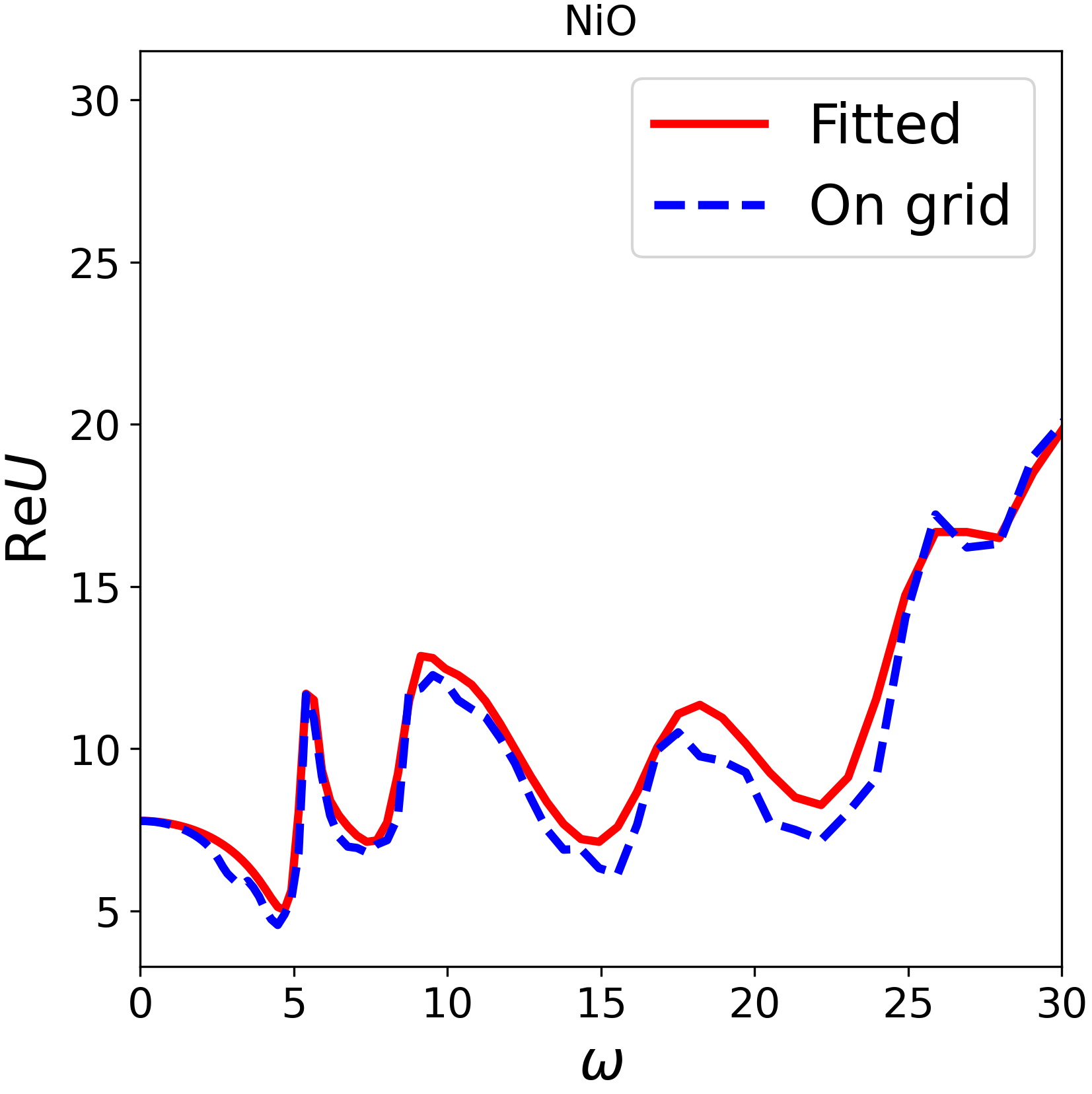}

    \caption{\small Low frequency range of the real part of $U(\omega)$ and SOP fitting for MnO, FeO, CoO and NiO.}
    \label{u_omega_fits_low_frequency_real}
\end{figure*}

\section{Projected density of states}
\begin{figure*}[!ht]    
    \includegraphics[width=0.49\textwidth]{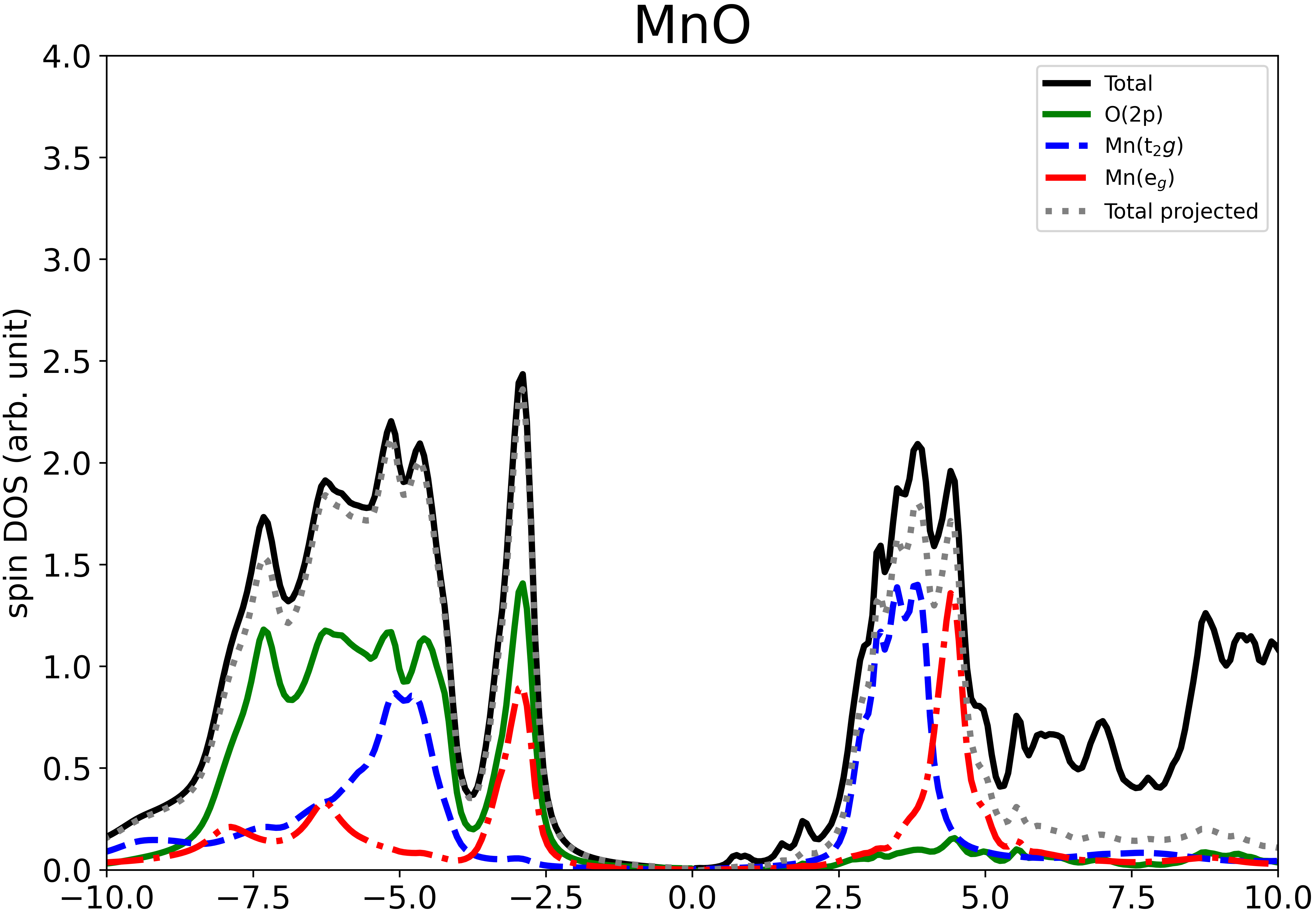}   \includegraphics[width=0.49\textwidth]{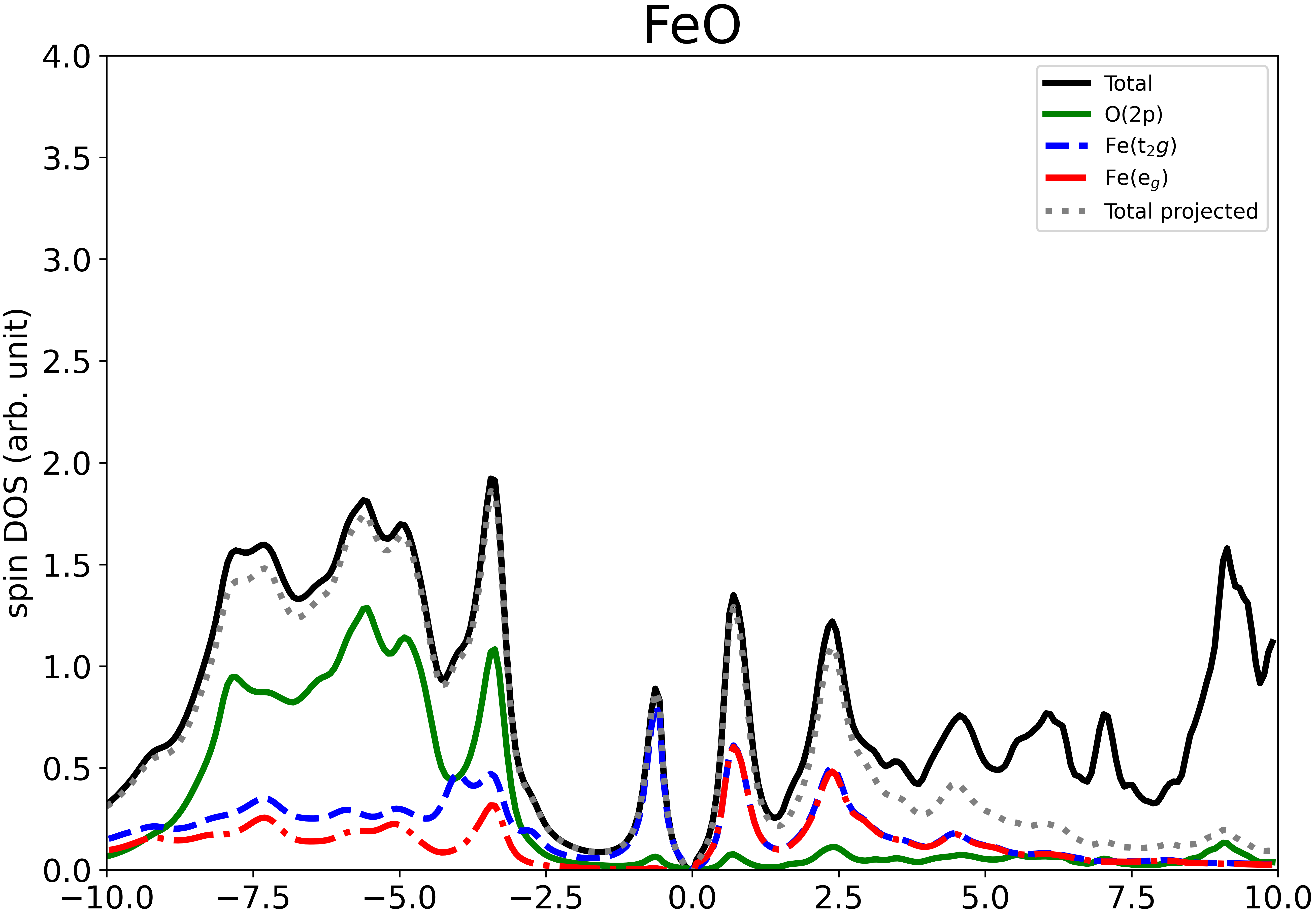}
    \includegraphics[width=0.49\textwidth]{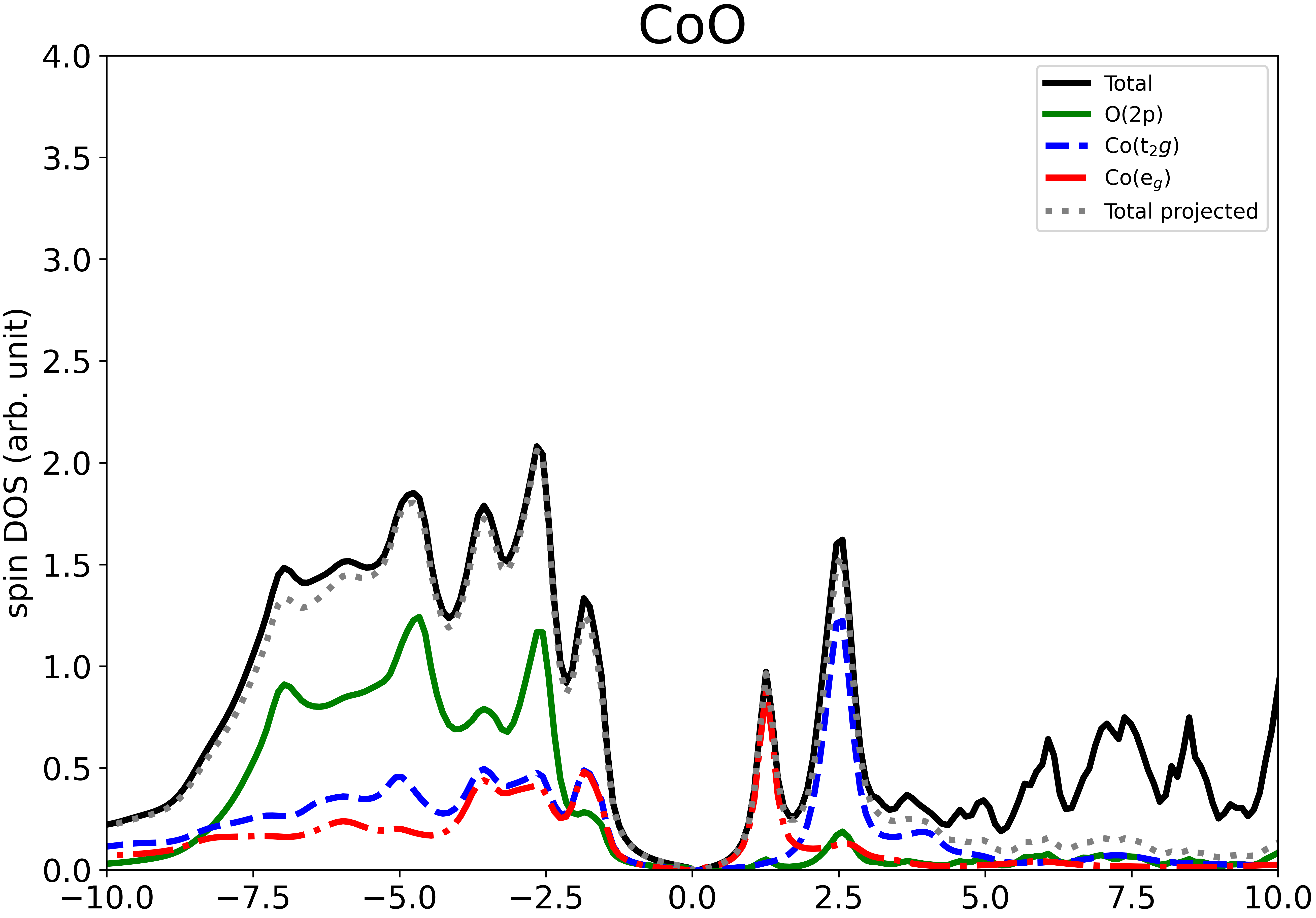}    
    \includegraphics[width=0.49\textwidth]{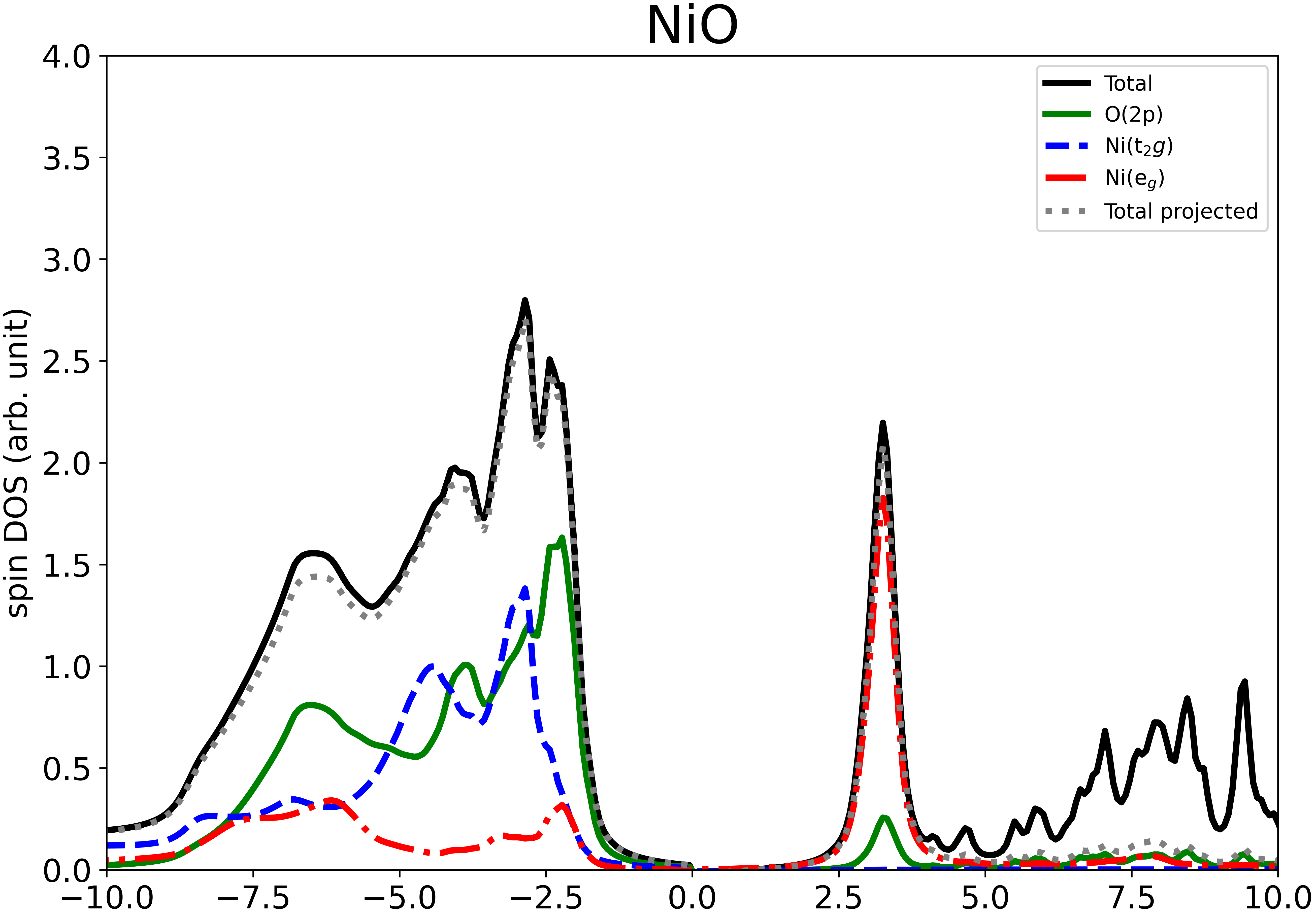}
      \caption{\small Projected spin density of states for DFT+dynH for MnO (a), FeO (b), CoO (c) and NiO (d). A broadening of 0.1 eV has used.}
    \label{proj_dosses}
\end{figure*}

In Fig.~\ref{proj_dosses} the projected density of states for the spin-channel are shown (with up and down channels being degenerate for AFM ground states). 
It can be seen that the four compounds have different Mott-Hubbard / charge-transfer characters, with FeO being the only case where the first excitation gap is entirely of Fe $3d$ nature. 
In NiO we can observe a sharp rise in the occupied spectrum which is very consistent with the experimental PES data~\cite{hufner_optical_1992}, see main text, and with an oxygen character of circa 60\% and Fe $t_{2g}$ and $e_g$ character for the remainer 40\% components of the peak, confirming the mixed Mott-Hubbard / charge-transfer nature of the compound.

\section{Computational details}
All DFT calculations are performed using the PWscf code of the \textsc{Quantum ESPRESSO} 
distribution~\cite{giannozzi_quantum_2009}.
The localized screened-potential $U(\omega)$ has been calculated using an in-house version of RESPACK~\cite{nakamura_respack_2021} modified to calculate collinear spin RPA polarization functions and spin Maximally Localized Wannier Functions matrix elements. 
The self-energy, Green's function, and spectral function are computed with a {\sc Python} dynamical Hubbard code~\cite{chiarotti_energies_2024}, here generalized to treat magnetism.
The pseudopotentials used for all the calculations are optimized norm-conserving pseudopotentials~\cite{hamann_optimized_2013} (PBEsol standard-precision, nc-sr-04) from the {\sc Pseudo Dojo} library~\cite{van_setten_pseudodojo_2018}.

We adopted a k-point mesh of 10x10x5 for DFT self-consistent cycles, with $200$ and $800$ Rydberg of wavefunction and charge energy cut-offs.
the RPA screening $W$ a grid of 6x6x3 q-points has been used, with $10$ Rydberg cut-off for the non-local field matrix element of the dieletric function, and $80$ bands expansion for the transition of the polarization function. 
The local screened interactions are calculated on $150$ frequencies and fitted with 23, 24, 24, 20 poles respectively for MnO, FeO, CoO and NiO.

The Wannierization window to build the MLWF starts $10$ eV below and goes to $10$ above the Fermi level of each, with 24 Wannier functions obtained with hydrogenoid $4s$, $3p$, $3d$ as initial guesses for the transition metals and $2p$ initial guesses for the oxygens.

The number of poles of the self-energy of Eq.~(1) in the main text is equal to the number of poles of the local Green's function times those of screened interaction $U(\omega)$. The latter brings typically an order $10$ of poles, while the former has poles at all Kohn-Sham states, which means that for a k-grid of 10x10x5 and 32 Kohn-Sham orbitals, one gets over $10000$ poles. In order to reduce the number of poles, i.e. the dimension of the AIM matrix to be diagonalized~\cite{chiarotti_energies_2024}, we employ a ``poles condensation'', please see Refs.~\cite{chiarotti_energies_2024,chiarotti_spectral_2023} for details. The number of poles used are respectively $736$, $648$, $840$ and $640$ for MnO, FeO, CoO and NiO.

\end{document}